\documentstyle[prl,aps,preprint,epsf]{revtex} 
\large  

\begin{document}
\draft

\title{Binary data corruption due to a Brownian agent} 
\author{T. J. Newman and Wannapong Triampo} 
\address{Department of Physics,\\ 
Virginia Polytechnic Institute and State University,\\ 
Blacksburg VA 24061, USA} 
\maketitle
\begin{abstract}
We introduce a model of binary data corruption
induced by a Brownian agent (active random walker) on a 
$d$-dimensional lattice. A continuum formulation allows the
exact calculation of several quantities related to the 
density of corrupted bits $\rho $ ; for example the
mean of $\rho$, and the density-density
correlation function. Excellent agreement is found
with the results from numerical simulations. We also
calculate the probability distribution of $\rho $
in $d=1$, which is found to be log-normal, indicating 
that the system is governed by extreme fluctuations.
\end{abstract}
\vspace{5mm} 

\newpage

\section{Introduction}

Brownian motion is one of the fundamental processes in Nature.
Originally observed in the irregular motion of pollen grains
by the botanist Brown\cite{brown}, and cast into the language
of the diffusion equation by Einstein\cite{einst}, it has now
been applied, in the mathematical framework of random walks\cite{hug}, 
to an enormous variety of processes in the physical sciences and beyond. 
A very rich field of research has been built up around the behavior of 
a random walk coupled to a disordered environment\cite{rwe}, a good
example being the anomalous diffusion of electrons in a disordered 
medium\cite{edis}. Also, one can consider a random walker {\it being} the
disordering agent in its environment. Applications of the latter
include the tagged diffusion of atoms in a crystal\cite{vmdr,hh,new},
or magnetic disordering due to a wandering vacancy\cite{schmitt}.

In this paper we will introduce a particularly simple example of an active 
random walker (or Brownian agent) disordering its environment. Although the 
model is interesting in its own right, we believe it will have a useful 
application to the study of data corruption in ultra-small storage devices.
Before pursuing this connection, we shall briefly describe the model
(which will be more carefully defined in the next section). The two
main features of the model are first, the Brownian agent (BA)
performs a pure random walk -- it is not affected by the
environment in any way. Second, the environment is bistable.
That is to say, it is composed of elements which may only exist
in one of two possible states (see Ref.\cite{hk} for a loosely 
related random walk process). Thus we can consider the environment
to be composed of binary data (our favored realization), magnetic
spins, chemical species A and B, and so on. As the BA wanders
through the environment it has a certain probability to switch
the value of an element in its immediate vicinity.
Thus, if we start with a system in which
all elements exist in the same state (`up' say), and introduce the
BA at the origin, then after some time, there will be a region
around the origin in which the elements will be found in
a mixture of `up' and `down' states. Naturally, the linear size
of the region will grow on average as $\sqrt{t}$. The more
subtle question concerns the degree of disordering which exists
for elements within this region, and also their spatial correlations..
As we shall see, the statistics of the disordered elements are
very rich. This is most convincingly demonstrated by the dominance
of extreme fluctuations; for instance, the distribution of
disordered elements is log-normal. Thus, typical and average
events are quite distinct, and become ever more so as time
proceeds.

Before giving an outline of the paper we shall say a few words
about the potential relevance of this system to data corruption.
With the advent of semiconductor memories (for dynamic random
access memories (DRAM's) and various types of read only
memories (ROM's)), there has been a tremendous drive within
the semiconductor industry to produce ever-smaller memory
devices\cite{mid,shar}. There are many properties ({\it e.g.} stability, 
power consumption, volatility, and cost) which must be balanced
in the design of such devices. These factors
determine the type of material used, and the geometry,
dimension and architecture of the device. (For instance, three
dimensional arrays have a very efficient address structure,
and are stable against interference from bombarding alpha 
particles, but are very expensive to produce\cite{mid}.)
One of the main issues is the stability, or reliability,
of the device. In semiconductor memories, there are many
physical effects which can create hard errors 
(destruction or corruption of the device itself) or
soft errors (corruption of the data stored in
the device). In the latter category, the most common
problems originate from electron clouds caused by
alpha particles, but soft errors may also arise from
electromigration and charge diffusion\cite{corr}. The key point
is that different corruption mechanisms operate on
different time scales (leading to the famous
bath-tub curve of device reliability\cite{shar}).
It is therefore important to know on what time-scales
one should expect significant corruption from a given
process. The model we propose here (namely
data corruption via a BA) is probably not relevant
for today's semiconductor devices, since there
are so many `mesoscopic' processes occurring on
the level of a flip-flop, that subtle correlations
due to a BA will be washed out. However, we can look ahead
to the new generation of (quantum) storage devices, 
in which a single electron (controlled in
a gate via coulomb blockade) can store one
bit of data. In this case, a microscopic BA may
indeed play an important role in data corruption, and
it will be necessary to understand its time-scales
and efficacy of operation, so that we can minimize its
influence. This paper constitutes a first step towards
gaining such an understanding.

The outline of this paper is as follows. In the next section
we shall carefully define the model (using discrete space
and time) within the master equation formulation of stochastic
processes. We shall derive some general statistical properties
of the process, but we shall not enter into any explicit
calculations. This is deferred to sections III-V in which we
introduce a very simple continuum theory for the process,
which is motivated by viewing the process as a stochastic
cellular automaton. In section III we derive this continuum 
theory and, using the complementary descriptions of quantum
mechanics ({\it i.e.} the Schr\"odinger equation and the
Feynman path integral), we shall demonstrate its equivalence
to the master equation formulation of section II. We then
examine the case of one spatial dimension in section IV.
The model is tractable, and a great deal of information 
may be derived concerning the mean density of disordered
elements, their spatial correlations, and finally 
their entire distribution function. In section V, we
briefly study higher dimensions, and derive some general
statistical properties for the process, for an arbitrary
spatial coupling between the BA and its environment. We also 
derive an expression for the mean density
of disordered elements in two dimensions. In section VI
we present results from extensive numerical simulations
of the discrete process. In all cases, we find good 
agreement between the simulation results and the 
predictions of the continuum theory.  We end the paper
with a summary of the work, and some ideas for future
study.

\section{Discrete formulation of the model}

We consider binary data bits on a $d$-dimensional hypercubic lattice.
For convenience we shall represent each bit by an Ising spin 
$\sigma _{\bf r}$ , where the index ${\bf r}$ represents a discrete 
lattice vector. The spin takes the
value $+1 (-1)$ for a data bit which is uncorrupted (corrupted). Thus,
the initial configuration is a lattice of spins, all of which take the
value $+1$. [We prefer to describe the system almost exclusively in terms
of the spin variables. Thus we shall use phrases such as `magnetization
density', or `global magnetization'. The translation of these quantities
to the corresponding properties for corrupted data bits is immediate
as one only need replace $\sigma _{\bf r}$ by $(1-2 n_{\bf r})$, where
$n_{\bf r}$ denotes the presence (with value unity) or absence (with
value zero) of a corrupted bit. Similarly, we shall often refer to the
average magnetization density $m$ , which is related to the average
density of corrupted bits ${\bar \rho}$ via $m=1-2{\bar \rho}$ .]
We denote the position of the BA by the lattice vector ${\bf R}$.
Each time step, the BA has a probability $p$ to make a jump to one of its
($2d$) nearest neighbors. For the sake of generality, we will not insist
that the BA always flips a spin ({\it i.e.} changes a data bit) as it moves.
Thus, on a given jump, we allow the BA to flip the spin
at the site it is leaving, with a probability $q$. 
In this section we shall describe the process via a 
master equation\cite{gard}. 
Namely, we shall define the dynamics through the evolution of the  
distribution $P({\bf R}, \lbrace \sigma _{\bf r} \rbrace , t)$, which is
the probability that at time $t$ the BA is at position ${\bf R}$, and the
spins have configuration $ \lbrace \sigma _{\bf r} \rbrace $.
Given the above rules, the master equation takes the form
\begin{eqnarray}
\label{master}
\nonumber
P({\bf R}, \lbrace \sigma _{\bf r} \rbrace , t+\delta t)  = 
(1-p) P({\bf R}, \lbrace \sigma _{\bf r} \rbrace , t) & + & {p(1-q)\over 2d}
\sum \limits _{\bf l}P({\bf R}+{\bf l}, \lbrace \sigma _{\bf r} \rbrace , t)\\
& + & {pq\over 2d}
\sum \limits _{\bf l}P({\bf R}+{\bf l}, \cdots, -\sigma _{{\bf R}+{\bf l}},
\cdots , t) \ ,
\end{eqnarray}
where $\lbrace {\bf l} \rbrace$ represent the $2d$ orthogonal lattice vectors
(which have magnitude $l$).

The natural quantities to extract from the distribution $P$ are conditional
averages. The simplest is the mean value of the spin at position ${\bf r}_{1}$
at time $t$, given the BA is at position ${\bf R}$. This is defined via
\begin{equation}
\label{lomag}
\Theta ({{\bf r}_{1}},t\ | \ {\bf R}) \equiv {\rm Tr} _{\sigma} \ 
\sigma _{{\bf r}_{1}} \ P({\bf R}, \lbrace \sigma _{\bf r} \rbrace, t) \ .
\end{equation}
Higher order conditional averages may be defined accordingly. Performing the
spin trace over the master equation with a weight of $\sigma _{\bf r}$ yields
\begin{eqnarray}
\label{lomageq}
\nonumber
\Theta ({\bf r},t+\delta t\ | \ {\bf R})  - 
\Theta ({\bf r},t \ | \ {\bf R})  & = & {p\over 2d}
\sum \limits _{\bf l} \bigl [ \Theta ({\bf r},t \ | \ {\bf R}+{\bf l}) - 
\Theta ({\bf r},t \ | \ {\bf R}) \bigr ]\\
& - & {pq\over d} \ \Theta ({\bf r},t \ | \ {\bf r}) 
\sum \limits _{\bf l} \delta _{{\bf r}, {\bf R}+{\bf l}} \ .
\end{eqnarray}
The above equation has a physically appealing form. The rate of change
of $\Theta $ has two contributions. The first is lattice diffusion, as given 
by the first sum on the r.h.s. The second contribution vanishes unless the
spin in question is in the immediate vicinity of the BA, in which case
it acts as a sink. 

At this point in the discussion it is worthwhile to consider the
continuum limit. Namely we take the time scale $\delta t$ and the 
lattice scale $l$ to zero, and define a diffusion constant $D \equiv
l^{2}p/\delta t$. We also introduce a coupling $\lambda \propto ql/\delta t$.
Then replacing the Kronecker $\delta$-function in Eq.(\ref{lomageq}) by
a Dirac $\delta$-function we find
\begin{equation}
\label{lomco}
\partial _{t}\Theta ({\bf r},t \ | \ {\bf R})  =
D\nabla _{\bf R}^{2} \Theta ({\bf r},t \ | \ {\bf R}) 
 - \lambda \ \Theta ({\bf r},t \ | \ {\bf r}) \delta ^{d} ({\bf r}-{\bf R}) \ .
\end{equation}
It is important to note that this continuum equation is not strictly derived
from Eq.(\ref{lomageq}), as we have not proved that the continuum limit
exists. In fact, we shall find that for $d \ge 2$, the lattice scale $l$
is crucial, and consequently we must soften the Dirac $\delta$-function
to a function $\Delta _{l}({\bf r})$ which is sharply peaked 
(over a linear scale $l$) with unit integral over ${\cal R}^{d}$. One 
actually expects this to be the case, as Eq.(\ref{lomco}) is the 
imaginary-time Schr\"odinger equation for a particle under the influence
of a repulsive  $\delta $-function potential. 
[Note the independent spatial variable in
this quantum system is ${\bf R}$, with the variable ${\bf r}$ simply
labelling the position of the potential.] 
It is well known\cite{land} that the repulsive $\delta$-function potential is
`invisible' to the particle for $d \ge 2$, and one usually cures this by
smearing the potential just as described above. The quantum mechanics
analogy will prove useful in the next section when we construct an
alternative continuum model.

Before leaving this section we shall indicate the derivation of a non-trivial
statistical relation hidden inside Eq.(\ref{lomageq}). First, we must
define the initial condition. As mentioned before, given that up-spins
denote uncorrupted data, the initial value of each spin is $+1$. There
is a slight subtlety of definition regarding the value of the spin at the site
where the BA is initially planted. (This position shall be taken to be the origin, without
any loss of generality.) We shall take this spin to be initially $-1$ so that
immediately after the BA has moved away the spin at the origin has value $+1$.
Thus we have 
\begin{equation}
\label{icp}
P({\bf R}, \lbrace \sigma _{\bf r} \rbrace, 0) = \delta _{{\bf R},{\bf 0}} 
\delta _{\sigma _{\bf 0},-1} \prod \limits _{{\bf r} \ne {\bf 0}} 
\delta _{\sigma _{\bf r},1} \ ,
\end{equation}
and consequently,
\begin{equation}
\label{ict}
\Theta ({\bf r}, 0 \ | \ {\bf R}) = \delta _{{\bf R},{\bf 0}} 
(1-2\delta _{{\bf r},{\bf 0}}) \ .
\end{equation}

We may obtain the following two averages from $\Theta ({\bf r},t \ | \ {\bf R})$. 
The first is the average value of the spin at the origin. This is simply
given by $m_{\bf 0}(t) = \sum _{\bf R} \Theta ({\bf 0},t \ | \ {\bf R})$. 
The second average
is the quantity $\sum _{\bf R} \Theta ({\bf R},t \ | \ {\bf R})$, which
corresponds to averaging the value of the spin at the site where the BA happens
to be at time $t$. One can prove that 
\begin{equation}
\label{iden}
\sum \limits _{\bf R} \Theta ({\bf 0},t \ | \ {\bf R}) =
\sum \limits _{\bf R} \Theta ({\bf R},t \ | \ {\bf R}) \ ,
\end{equation}
for all $t$. We arrive at the above result by essentially solving
the partial difference equation (\ref{lomageq}) using discrete
Fourier and Laplace transforms. The details can be found in
Appendix A. 

This result is useful for proving a more physically relevant
relation. Let us denote the average global magnetization by 
\begin{equation}
\label{glomag}
M(t) = \sum \limits _{\bf R} \sum \limits _{\bf r} 
[ \Theta ({\bf r},0 \ | \ {\bf R}) - 
\Theta ({\bf r},t \ | \ {\bf R})] \ ,
\end{equation}
where we have defined it relative to the (infinite) initial 
magnetization. This quantity essentially measures the 
average of the total number of corrupted bits (up to a
factor of 2). Summing Eq.(\ref{lomageq}) over ${\bf r}$ and 
${\bf R}$ gives
\begin{equation}
\label{glomag2}
M(t+\delta t) - M(t) = 2pq \sum \limits _{\bf r} 
\Theta ({\bf r},t \ | \ {\bf r}) \ . 
\end{equation}
Then using (\ref{iden}) we can rewrite the above relation in
the form
\begin{equation}
\label{glomag3}
M(t+\delta t) - M(t) = 2pq \sum \limits _{\bf R} 
\Theta ({\bf 0},t \ | \ {\bf R}) = 2pq \ m_{\bf 0}(t) \ .
\end{equation}
In other words, the rate of change of the mean global
magnetization is proportional to the mean magnetization
density at the origin. This is a non-trivial relation
between a global and a local quantity.

In principle, one can obtain exact results for many 
interesting quantities (like the mean magnetization
density, or correlation functions) by directly solving
for the conditional averages, as illustrated in 
Appendix A. However, we prefer to obtain results
from a continuum theory; partly because the calculations
are a little easier, but more importantly because we can access
more sophisticated properties of the system, such as
the probability distribution of the coarse-grained
magnetization density.

\section{Continuum theory}

In this section we shall motivate a particularly simple
continuum description of the data corruption process,
and show its equivalence to the discrete theory of the
previous section.

There is an alternative method of characterizing the evolution
of the system, other than using the evolution of the 
probability distribution  
$P({\bf R}, \lbrace \sigma _{\bf r} \rbrace , t)$
via the master equation. This method consists of writing
the local rules for the process in the spirit of a
stochastic cellular automaton (SCA)\cite{wolf}. Let us focus on
the case that at each time step the BA makes a 
random jump to one of its nearest neighbors, and
that the spin at the site which it leaves behind,
definitely flips. This corresponds to setting $p=q=1$.
The local rules for such a process are easily written
down. Let us denote the time-dependent position of
the BA by ${\bf R}(t)$, a randomly chosen unit lattice
vector by ${\bf l}(t)$, and the time-dependent value
of the spin at site ${\bf r}$ by $\sigma _{\bf r}(t)$. 
Then we have
\begin{eqnarray}
\label{scadis1}
{\bf R}(t+\delta t) & = & {\bf R}(t) + {\bf l}(t) \\
\label{scadis2}
\sigma _{\bf r}(t+\delta t) & = & \sigma _{\bf r}(t)
(1-2\delta _{{\bf r},{\bf R}(t)})  
\end{eqnarray}

We are interested in a continuum limit of these two rules.
The first is nothing more
than a random walk. We take the lattice vector ${\bf R}(t)$
to be a continuum vector ({\it i.e.} each of the $d$ components 
is a real number), and we replace the random unit lattice
vector ${\bf l}(t)$ by a continuum vector ${\bf {\xi}} (t)$, each
component of which is a uncorrelated Gaussian random variable with zero 
mean ({\it i.e.} $\xi _{i} (t)$ is a white noise process). 
The correlator of ${\bf {\xi}}$ is given by 
\begin{equation}
\label{corrxi}
\langle \xi _{i} (t) \xi _{j}(t') \rangle = D'\delta _{i,j}\delta (t-t') \ ,
\end{equation}
where here and henceforth, angled brackets indicate an average
over the noise (or equivalently the paths of the BA).
Then, on taking $\delta t \rightarrow 0$, Eq.(\ref{scadis1}) assumes the form
\begin{equation}
\label{scacont1}
{d {\bf R} \over dt} = {\bf {\xi }}(t) \ ,
\end{equation}
which is the familiar equation for a continuum random walker where $D'$
is the diffusion constant\cite{gard}. 
The second SCA rule is more complicated to generalize to the
continuum. As a first step let us define a coarse-grained
magnetization density $\phi ({\bf r},t)$ in the following way.
We imagine defining a large region around the lattice
site ${\bf r}$ and summing all the spins in that region.
Their sum (suitably normalized) constitutes $\phi ({\bf r},t)$,
with the label ${\bf r}$ denoting a point in the ${\cal R}^{d}$
continuum. An entirely analogous procedure is used in
motivating the Landau-Wilson free energy functional from
the Ising model of ferromagnetism\cite{crit}. The difficulty in our
case is that we cannot derive a closed equation for $\phi $
from the discrete rule (\ref{scadis2}). We therefore make
the following approximation. Splitting the r.h.s. of  
(\ref{scadis2}) into two pieces, we see that the first may
be taken over to the l.h.s. which may then be
taken to be a time derivative in the limit of 
$\delta t \rightarrow 0$ . The second piece resembles
a decay term centered at ${\bf r}={\bf R}$. So, we postulate
that the coarse-grained magnetization density satisfies 
\begin{equation}
\label{scacont2}
\partial _{t} \phi ({\bf r},t) = -\lambda ' \phi ({\bf r},t)
\delta ^{d} ({\bf r}-{\bf R}(t)) \ ,
\end{equation}
where $\lambda '$ is a phenomenological parameter which
describes how strongly the magnetization density is
coupled to the BA. We stress that the field $\phi ({\bf r},t)$
is a function of the continuous space and time variables
${\bf r}$ and $t$, and a {\it functional} of the path ${\bf R}(t)$
of the BA.

Now, the above heuristic derivation of the continuum theory
was based on a SCA for the case in which the BA always
moves ($p=1$), and for which the spin located at the 
previous BA position is always flipped ($q=1$). In
general $p$ and $q$ are both less than unity. Intuitively
we expect a very simple renormalization of our
phenomenological parameters as $p$ and $q$ are changed.
The diffusion constant $D'$ should be proportional to $p$,
and the strength of the spin-BA coupling $\lambda '$ should be
proportional to both $p$ and, more importantly, $q$.
Thus we see a very close correspondence between $D'$ and
$\lambda '$ in the current continuum theory, and the
parameters $D$ and $\lambda $ which were introduced 
in the continuum limit (\ref{lomco}) of the discrete
equation (\ref{lomageq}). In fact they are identical,
as will emerge in the following discussion.

One of the positive features of the continuum theory as
described by (\ref{scacont2}) is that one may 
immediately integrate the equation to find the
magnetization density as an explicit functional
of the path of the BA. As an initial condition we
take $\phi ({\bf r},0) = 1$ for all ${\bf r}$. The
subtlety encountered in the discrete theory concerning
the initial value of the spin at the origin disappears
here since the coarse grained function $\phi $ is
not sensitive to the value of one inverted spin.
Straightforward integration of (\ref{scacont2}) yields
\begin{equation}
\label{solcon}
\phi ({\bf r},t) = \exp \left [ -\lambda ' \int \limits
_{0}^{t} dt' \ \delta ^{d} ({\bf r}-{\bf R}(t')) \right ] \ . 
\end{equation}
It is important to note at this stage that the magnetization
density $\phi $ is clearly positive for all ${\bf r}$ and $t$.
Therefore within our continuum formulation, we have 
ignored paths of the BA which create large
patches containing a majority of down spins ({\it i.e.} 
corrupted bits). Such patches
will occur, but their frequency of occurrence
is certainly very small since the system starts in
a completely uncorrupted state. For instance, the probability for
the BA to create a purely negative domain of $N$ spins 
is of the order $e^{-N}$. Therefore, so long
as we coarse-grain the original spin model over a
sufficiently large scale, we can be confident that the
most important configurations have been retained in the
continuum theory. Ultimately, one must justify such
an approximation {\it a posteriori} by comparison
with either results from the discrete theory, or
from numerical simulations. As we shall see, both
of these support the current continuum model and the
approximations contained therein.

We shall now connect the
continuum theory as given by (\ref{solcon}), and 
the continuum limit (\ref{lomco}) of the discrete theory, as
given by (\ref{lomageq}). The mean magnetization
density in the discrete theory is given by
\begin{equation}
\label{mmd1}
m_{\bf r}(t) = \sum \limits _{\bf R} \Theta ({\bf r},t \ | {\bf R}) 
{{\rm CL}\atop \longrightarrow } \int d^{d}R \ 
\Theta ({\bf r},t \ | {\bf R})
\end{equation}
where in the continuum limit, we have replaced the sum over
BA positions by an integral, and the field $\Theta $
satisfies the imaginary-time Schr\"odinger equation
as given in (\ref{lomco}). In the alternative continuum
theory, we can find the mean magnetization density by
averaging the coarse-grained density $\phi ({\bf r}, t)$
over all paths ${\bf R}(t)$. Each path is weighted
by a Gaussian factor 
\begin{equation}
\label{weight}
{\cal N} {\rm exp} \left \lbrace -{1 \over 2D'} \int \limits
_{0}^{t} dt' \ {\bf {\xi }}(t')^{2} \right \rbrace
=
{\cal N} {\rm exp} \left \lbrace -{1 \over 2D'} \int \limits
_{0}^{t} dt' \ \left ( {d{\bf R}\over dt'} \right )^{2} 
\right \rbrace \ ,
\end{equation}
where ${\cal N}$ is a normalization factor.
Therefore we can write the average of $\phi $ as a 
functional integral
\begin{eqnarray}
\label{mmd2}
\nonumber
m({\bf r},t)=\langle \phi ({\bf r},t) \rangle & = &
{\cal N} \int {\cal D}{\bf R}(t') \ 
{\rm exp} \left \lbrace -{1 \over 2D'} \int \limits
_{0}^{t} dt' \ \left ( {d{\bf R}\over dt'} \right )^{2} 
\right \rbrace \ \phi ({\bf r},t) \ , \\
\nonumber
& = & {\cal N} \int {\cal D}{\bf R}(t') \ 
{\rm exp} \left \lbrace  - \int \limits
_{0}^{t} dt' \ \left [ {1\over 2D'}
\left ( {d{\bf R}\over dt'} \right )^{2} 
+ \lambda ' \delta ^{d} ({\bf r}-{\bf R}(t'))
\right ] \right \rbrace \ , \\
& = & {\cal N} \int d^{d}R_{f} \int \limits
^{{\bf R}_{f}} {\cal D}{\bf R}(t') \ 
{\rm exp} \left \lbrace  - \int \limits
_{0}^{t} dt' \ \left [ {1\over 2D'}
\left ( {d{\bf R}\over dt'} \right )^{2} 
+ \lambda ' \delta ^{d} ({\bf r}-{\bf R}(t'))
\right ] \right \rbrace \ , 
\end{eqnarray}
where we have used (\ref{solcon}) in going from
the first line to the second, and we have introduced
the final position of the BA ({\it i.e.} {\bf R}(t)) 
as a free integration
variable ${\bf R}_{f}$ in rewriting the second line
as the third. The reason for this cosmetic change
is to make explicit the fact that 
$\langle \phi ({\bf r},t) \rangle$ can be expressed
as a spatial integral over the final BA position,
where the integrand is itself a path integral over
BA trajectories.
This path integral is nothing more than a re-expression
of the solution of an imaginary time Schr\"odinger
equation (using the well-known Feynman path integral
formulation of quantum mechanics\cite{fh}) for a particle
in a repulsive $\delta$-function potential.
We can now see the connection: Eq. (\ref{mmd2}) is
an exact restatement of Eqs. (\ref{lomco}) and
(\ref{mmd1}), with the identification  
$D'=D$ and $\lambda ' = \lambda $. [So, henceforth
we shall drop the primes in the material parameters.]
To summarize, by utilizing the complementary formulations
of quantum mechanics via the Schr\"odinger equation and the
Feynman path integral, we have shown that the continuum
limit of the master equation is identical to the
continuum theory constructed at the beginning of this
section. 

Two final points are in order. First, as noted in
the previous section, the $\delta$-function
potential must be replaced by a smeared function
$\Delta _{l}({\bf r})$ for $d \ge 2$; thus
in our continuum theory encapsulated in 
Eq.(\ref{scacont2}), we shall make a similar
replacement when studying two or higher dimensions.
Second, we have established a connection between
the master equation and Eq.(\ref{scacont2})
only at the level of the first moment. It is
straightforward to extend each formulation
to higher order correlation functions, and indeed
one finds an exact correspondence. For instance,
we can define the conditional spin-spin correlation function
within the discrete theory
\begin{equation}
\label{spincor}
\Theta ({{\bf r}_{1}},{{\bf r}_{2}},t\ | \ {\bf R}) 
\equiv {\rm Tr} _{\sigma} \ 
\sigma _{{\bf r}_{1}} \sigma _{{\bf r}_{2}} \ 
P({\bf R}, \lbrace \sigma _{\bf r} \rbrace, t) \ .
\end{equation}
Using the master equation one can show that in the continuum limit
this function satisfies the Schr\"odinger equation for a particle
under the influence of two repulsive $\delta $-function
potentials located at ${\bf r}_{1}$ and ${\bf r}_{2}$. Similarly,
we can construct the coarse-grained two-point correlation 
function from Eq.(\ref{solcon}) by evaluating
$\langle \phi ({\bf r}_{1},t)\phi ({\bf r}_{2},t) \rangle$.
It is easy to see that this quantity is given by an integral
over the analogous path-integral for two repulsive 
$\delta$-function potentials. 

Having completed our formulation of a simple continuum theory,
and shown its equivalence to the continuum limit of the
master equation, we shall proceed to the next section in
which we present a comprehensive solution of the model
in one dimension.

\section{Results in one dimension}

In this section we restrict ourselves to one dimension. This does
not necessarily mean a single chain of sites. Rather, we shall 
exclusively study the continuum theory of the last section, and
in this case, for large enough times, $d=1$ refers to any system
which has an infinite longitudinal dimension, and finite transverse 
dimensions (for instance an infinitely long strip). This is the
case, since as time proceeds, the correlation length 
will eventually become greater than the transverse size of the 
system, thereby only allowing the longitudinal fluctuations 
to continue growing, as is the case in a strictly one-dimensional system.

The continuum model described in the previous section can be
viewed as a `non-conserved' cousin of the continuum theory
of vacancy mediated diffusion (a process which in the 
spin language conserves magnetization) introduced recently\cite{new}.
An exact analysis of the latter theory was possible using infinite
order perturbation theory in the spin-BA coupling $\lambda $. We
shall use the same technique here, as it leads rather directly to
a full solution. Alternatively, one may solve the Schr\"odinger
equation for the conditional averages. However, there are some
important quantities (like the distribution of the magnetization
density) which cannot be easily recovered from the latter approach.

Our starting point is the integrated solution of the continuum
formulation as given in Eq.(\ref{solcon}). First, we shall derive an
expression for the magnetization density $m(x,t)$. Performing
a direct average of Eq.(\ref{solcon}) and expanding in powers of
$\lambda \ (=\lambda ')$, we have
\begin{equation}
\label{avden}
m(x,t) = \langle \phi (x,t) \rangle = \sum \limits _{n=0}
^{\infty} (-\lambda )^{n} \chi _{n}(x,t) \ ,
\end{equation}
where $\chi _{0}(x,t)=1$, and for $n>0$
\begin{equation}
\label{defchi}
\chi _{n}(x,t) = {1\over n!} \left \langle \left [
\int \limits _{0}^{t} d\tau \ \delta (x-R(\tau)) \right ]^{n}
\right \rangle \ . 
\end{equation}
We refer the reader to Appendix B in which the above average
is explicitly calculated. The result is
\begin{equation}
\label{solchi}
\chi _{n}(x,t) = \int \limits _{0}^{t} d\tau _{1} 
\int \limits _{0}^{\tau _{1}} d\tau _{2} \cdots 
\int \limits _{0}^{\tau _{n-1}} d\tau _{n} \ 
g(0,\tau _{1}-\tau _{2}) \cdots
g(0,\tau _{n-1}-\tau _{n}) g(x,\tau _{n}) \ ,  
\end{equation}
where $g(x,t)=(2\pi Dt)^{-1/2}\exp (-x^{2}/2Dt)$ is the probability 
density of the BA.

The structure of Eq.(\ref{solchi}) is that of an $n$-fold convolution,
so we may utilize a Laplace transform to good effect. We have (for $n>0$)
\begin{equation}
\label{ltchi}
{\hat \chi}_{n}(x,s) \equiv \int \limits _{0}^{\infty } dt \
e^{-st} \ \chi _{n}(x,t) = {1 \over s} \ {\hat g}(0,s)^{n-1} {\hat g}(x,s) \ .
\end{equation}
Performing the sum over these functions as dictated by Eq.(\ref{avden}) 
we find
\begin{equation}
\label{ltden}
{\hat m}(x,s) = {1 \over s} \left [ 1 - {\lambda {\hat g}(x,s) \over
1+\lambda {\hat g}(0,s)} \right ] \ .
\end{equation}
We note in passing that a similar result is easily derived for any
$d \in (0,2)$. The case of $d \ge 2$ is more complicated as the function
$g(0,t)$ is no longer integrable.

This expression for the Laplace transform of $m$ is exact. This will prove
to be important when we come to evaluate the distribution function of 
${\cal P}(m)$. The inverse of the Laplace transform is given by
\begin{equation}
\label{iltden}
m(x,t) = {\rm erf} \left [ {|x|\over (2Dt)^{1/2}} \right ]
+ \exp \left ( {\lambda |x| \over D} + {\lambda ^{2}t \over 2D} \right )
\ {\rm erfc} \left [ \lambda \left ( {t \over 2D} \right )^{1/2} + 
{|x|\over (2Dt)^{1/2}} \right ] \ ,  
\end{equation}
where ${\rm erf}(z)$ and ${\rm erfc}(z)$ are error functions\cite{as}.
Considering the long time behavior of the above expression, we have
for $x=0$
\begin{equation}
\label{resden0}
m(0,t) = \left ( {2D \over \pi \lambda ^{2} t} \right )^{1/2}
\left [ 1 + O \left ( {D \over \lambda ^{2}t} \right )
\right ] \ . 
\end{equation}
One can also retrieve the spatial behavior with little effort.
For small $x$ we have
\begin{equation}
\label{resdensx1}
m(x,t) = m(0,t) +
\left ( {2 x^{2} \over \pi Dt} \right )^{1/2} + \cdots \ \ \ \ , 
\ x \ll (Dt)^{1/2} \ .
\end{equation}
The large $x$ behavior has two regimes:
\begin{eqnarray}
\label{resdensx2}
m(x,t) & = & 1 - \left ( {2Dt \over \pi x^{2}} \right ) ^{1/2} \exp 
\left ( {-x^{2} \over 2Dt } \right ) + \cdots \ \ \ \ , 
(Dt)^{1/2} \ll x \ll \lambda t \\
m(x,t) & = & 1 - \left ( {\lambda t \over |x| } \right ) 
\left ( {2Dt \over \pi x^{2}} \right ) ^{1/2} \exp 
\left ( {-x^{2} \over 2Dt } \right ) + \cdots \ \ \ \ , x \gg \lambda t \ .
\end{eqnarray}
It is interesting to note that apart from the natural diffusion scale
$(Dt)^{1/2}$, there is a larger `ballistic' scale $\lambda t$ in the
system, beyond which the disordering efficacy of the BA is much reduced,
since it makes so few visits to these distant sites.
There is no simple ({\i.e.} single length) scaling form for $m(x,t)$.

Next we consider the continuum analogues of Eqs.(\ref{iden})-(\ref{glomag3}).
We define the average global magnetization (relative to its initial value) as
\begin{equation}
\label{glomagc1}
M(t) = \int \limits _{-\infty}^{\infty} dx \ \left [
\langle \phi (x,0) \rangle - \langle \phi (x,t) \rangle \right ] \ ,
\end{equation}
which may be compared to the discrete version in Eq.(\ref{glomag}).
Integrating and averaging the continuum model (\ref{scacont2})
yields
\begin{equation}
\label{glomagc2}
{dM(t) \over dt} = \lambda \langle \phi (R(t),t) \rangle \ ,
\end{equation} 
which is to be compared with Eq.(\ref{glomag2}). 
This last equation indicates that we may explicitly find an expression
for $\langle \phi (R(t),t) \rangle$ by calculating the time derivative
of the spatial integral of $m(x,t)=\langle \phi (x,t) \rangle$. This may be
done at the level of the perturbation series (\ref{avden}), from which one
may show that the following relation holds exactly, for all times:
\begin{equation}
\label{idenc}
m(0,t) = \langle \phi (R(t),t) \rangle \ ,
\end{equation} 
which is the continuum analogue of Eq.(\ref{iden}). Finally, combining
Eqs.(\ref{glomagc2}) and (\ref{idenc}) we have
\begin{equation}
\label{glomagc3}
{dM(t) \over dt} = \lambda m(0,t) \ .
\end{equation} 
Thus, the non-trivial relation between the rate of change of the 
global magnetization, and the mean of the magnetization density at the
origin, is seen to be exact within the continuum model (which complements
the exact relation $(\ref{glomag3})$ found in the discrete framework).
Similar results are easily obtained for all $d \in (0,2)$. In section V
we shall derive a slightly more complicated form of $(\ref{glomagc3})$ 
(involving the smearing function $\Delta _{l}$) 
which is appropriate for higher dimensional systems.
Directly from Eqs.(\ref{resden0}) and (\ref{glomagc3}) we note
that $M(t) \sim \sqrt{t} $ ; thus, the average number of corrupted bits in
$d=1$ increases as the square root of time.

We now turn to spatial correlations in the system. These are most easily
probed via the two-point correlation function
\begin{eqnarray}
\label{corrfnc}
\nonumber
C(x,t) & = & \langle \phi (x,t) \phi (0,t) \rangle \\
\nonumber
\\
& = & \left \langle \exp \left (  -\lambda \int \limits _{0}^{t} dt' \ 
\bigl [ \delta (x-R(t')) + \delta (R(t')) \bigr ] \right ) \right
\rangle \ ,
\end{eqnarray}
where we have used the solution (\ref{solcon}) in the second line.
This average can be calculated using infinite order perturbation
theory in $\lambda $, just as was used to evaluate $m(x,t)$. We write 
\begin{equation}
\label{pertexpc}
C(x,t) = \sum \limits _{n=0}^{\infty} (-\lambda )^{n}c_{n}(x,t) \ ,
\end{equation}
with $c_{0}=1$. For $n>0$, a given term $c_{n}$ can be explicitly evaluated by
making integral representations of the $n$ $\delta$-functions, and performing
the average over the paths $R(t)$ (as described in Appendix B). Thus one has
\begin{eqnarray}
\label{corrcn}
\nonumber
c_{n}(x,t) =  \int \limits _{0}^{t} d\tau _{1} 
\int \limits _{0}^{\tau _{1}} d\tau _{2} \cdots 
\int \limits _{0}^{\tau _{n-1}} d\tau _{n} & &   
\int {dk_{1}\over 2\pi} \ (1+e^{-ik_{1}x}) \cdots
\int {dk_{n}\over 2\pi} \ (1+e^{-ik_{n}x}) \\
& \times & \exp \left [ -{D\over 2} 
\sum \limits _{m=1}^{n} \left (
\sum \limits _{l=1}^{m} k_{l}^{2} \right )
(\tau _{m}-\tau _{m+1}) \right ] \ , 
\end{eqnarray}
with the notation $\tau _{n+1} \equiv 0$.
This $2n$-fold integral can be reduced using Laplace transform in time,
such that the integrals over $\lbrace k_{i} \rbrace $ may be performed,
as described in Appendix C. The result is
\begin{equation}
\label{ltcorrn}
{\hat c}_{n}(x,s) = {1 \over s (2sD)^{n/2} } \left (
1+\exp \left [ - \left ( {2s \over D} \right )^{1/2} |x| \right ]
\right )^{n} \ .
\end{equation}
Summing over these functions with a weight of $(-\lambda )^{n}$ and 
inverting the Laplace transform using a Bromwich integral\cite{jj}, we have 
\begin{equation}
\label{fcorrfn}
C(x,t) = \int _{\gamma } {ds \over 2\pi i} \ {\exp (s{\tilde t}) \over
s^{1/2} [1+s^{1/2} + \exp (-s^{1/2}|{\tilde x}|) ] } \ , 
\end{equation}
where, as usual, the contour $\gamma $ is parallel to the imaginary axis, 
and to the right of any singularities. We have rescaled space and time 
as ${\tilde x}=x\lambda /D$ and ${\tilde t}=t\lambda ^{2}/2D$.

This integral may be evaluated for large ${\tilde t}$ in the following 
way. We re-express the integral as an expansion in powers of
$\exp (-s^{1/2}|{\tilde x}|)$ (which is {\it not} the same as our original
expansion in powers of $\lambda $). So we have
\begin{equation}
\label{corrfnexp2}
C(x,t) = \sum _{n=0}^{\infty} (-1)^{n}I_{n}({\tilde x},{\tilde t}) \ ,
\end{equation}
with
\begin{eqnarray}
\label{corrints}
\nonumber
I_{n} & = & \int _{\gamma } {ds \over 2\pi i} \ \
{e^{s{\tilde t}-ns^{1/2}|{\tilde x}|} \over
s^{1/2} (1+s^{1/2})^{n+1} } \\
\nonumber
\\
& = & {1\over 2\pi } \int \limits _{0}^{\infty} du \ {e^{-u{\tilde t}}
\over u^{1/2}} \ \left [ {e^{-inu^{1/2}|{\tilde x}|} \over (1+iu^{1/2})^{n}}
+ {e^{inu^{1/2}|{\tilde x}|} \over (1-iu^{1/2})^{n}} \right ] \ ,
\end{eqnarray}
where the second line is the explicit form of the integral after 
integrating around the only singularity -- a branch point located at 
$s=0$. The integral over $u$ may be simplified for ${\tilde t} \gg 1$ to
give
\begin{eqnarray}
\label{morecorrints}
\nonumber
I_{n} & \simeq & {1\over \pi {\tilde t}^{1/2} } \int \limits _{0}^{\infty} 
du \ {e^{-u}\over u^{1/2}} \cos \left ( {nu^{1/2}|{\tilde x}| \over 
{\tilde t}^{1/2} } \right ) \\
& = & {1 \over (\pi {\tilde t})^{1/2}} \exp \left (-{n^{2}{\tilde x}^{2}
\over 4{\tilde t}} \right ) \ .
\end{eqnarray}
In these rescaled units, we have from Eq.(\ref{resden0}) $m(0,{\tilde t})
\simeq (\pi {\tilde t})^{-1/2}$ (for large ${\tilde t}$). Thus, we may
resum the functions $\lbrace I_{n} \rbrace $ to find
\begin{eqnarray}
\label{corrfinal}
\nonumber
{C(x,t) \over m(0,t)} \ & \simeq & \ \sum \limits _{n=0}^{\infty}
(-1)^{n} \exp \left (-{n^{2}{\tilde x}^{2} \over 4{\tilde t}} \right ) \\
\nonumber
\\
& = & \ {1\over 2} \left [ 1 + \theta _{4} \left ( 
0,  \exp (-x^{2}/2Dt) \right ) \right ] \ ,
\end{eqnarray}
where $\theta _{4}(q,z)$ is a Jacobi theta function (with norm $q$)\cite{as}.
Note, we have written the last line in unscaled variables, and we see that
the ratio of the correlation function to $m(0,t)$ does not depend on 
$\lambda $ for large times. The behavior of $C(x,t)$ in the limits of 
large and small $x$ are as follows. For large $x$, the fields at the origin
and at $x$ will be uncorrelated, so that $C(x,t) \simeq m(0,t)m(x,t)
\simeq m(0,t)$, the latter result following since $m(x,t) \simeq 1$ for
$x \gg \sqrt {t}$. At the other extreme, as $x \rightarrow 0$, 
$C(x,t) \rightarrow \langle \phi (0,t) ^{2} \rangle$. Referring to the 
exact solution of the continuum model, Eq.(\ref{solcon}), one can see that
the second moment of the magnetization density is actually given exactly 
by $m(0,t;2\lambda )$ (where the last (optional) argument
indicates the parametric dependence on the spin-BA coupling). 
So for long times we take the 
expression for $m(0,t;\lambda)$ given in Eq.(\ref{resden0}) and replace $\lambda$
by $2\lambda $. Therefore $ \langle \phi (0,t) ^{2} \rangle \simeq
m(0,t;\lambda)/2$ for ${\tilde t} \gg 1$. 
Thus, the limits of the function $C(x,t)/m(0,t)$
are 1/2 (for small $x$) and unity (for large $x$), which is naturally
consistent with the analytic form given above in terms of the Jacobi
theta function. In section VI we shall compare this  
expression with results from a numerical simulation of the discrete model
described earlier.

To complete our study of the properties of this system in one dimension,
we shall consider the complete probability distribution 
${\cal P}(\phi, x, t)$ of the magnetization
density. We shall be able to calculate this exactly, since i) we can see
from (\ref{solcon}) that the $n^{\rm th}$ moment of the density is related
to the mean density with a replacement $\lambda \rightarrow n\lambda $; and
ii) we have an {\it exact} expression for the mean density (albeit in the 
Laplace transform variable $s$). The first point is a fortuitous property
of our continuum model which we should certainly exploit. The second
property is less obvious. One might imagine that, given we know all density
moments via the first, even the asymptotic form for the mean density 
would be sufficient to calculate the probability distribution (for large 
times). This is not the case as we shall see -- the complete analytic 
structure of $m(x,t;\lambda)$ is required in order to reconstruct the 
distribution ${\cal P}$.

We define ${\cal P}$ via
\begin{equation}
\label{probden}
{\cal P}(\phi, x, t) = \langle \delta (\phi - \phi _{R}(x,t)) \rangle \ ,
\end{equation}
where $\phi _{R}(x,t)$ is the stochastic field solution given in Eq.
(\ref{solcon}). We can re-express the $\delta $-function using a frequency
integral, and expand in powers of the field as follows:
\begin{eqnarray}
\label{deltaex}
\nonumber
{\cal P}(\phi , x, t) & = & \int \limits _{-\infty}^{\infty} 
{d\omega \over 2\pi}
\ e^{-i\omega \phi} \ \langle \exp \left [i\omega 
\phi _{R}(x,t) \right ] \rangle \\
\nonumber
& = &  \int \limits _{-\infty}^{\infty} {d\omega \over 2\pi}
\ e^{-i\omega \phi } \ \sum \limits _{n=0}^{\infty} {(i\omega )^{n}\over
n!} \langle \phi _{R}(x,t)^{n} \rangle \\
& = &  \int \limits _{-\infty}^{\infty} {d\omega \over 2\pi}
\ e^{-i\omega \phi } \ \sum \limits _{n=0}^{\infty} {(i\omega )^{n}\over
n!} m(x,t; n\lambda ) \ ,
\end{eqnarray}
the last line following from the property i) alluded to above.

So the Laplace transform (over time) of ${\cal P}$ is given in terms of the 
Laplace transform of $m(x,t;n\lambda) $. From Eq.(\ref{ltden}) we have
\begin{eqnarray}
\label{ltdengen}
\nonumber
{\hat m}(x,s;n\lambda) & = & {1 \over s} \left [ 1 - 
{n\lambda {\hat g}(x,s) \over 1+n\lambda {\hat g}(o,s)} \right ] \\
\nonumber
\\
& = & {{\hat g}(0,s)-{\hat g}(x,s) \over s{\hat g}(0,s)} \ + 
\ {{\hat g}(x,s) \over s{\hat g}(0,s)(1+n\lambda {\hat g}(0,s))} \ ,
\end{eqnarray}
where the second line follows from some algebraic manipulations.
The first term is easily handled as it is independent of $n$. Thus the
sum over $n$ for this term (as is required in Eq.(\ref{deltaex})) yields a factor of
$\exp (i\omega)$ which finally yields a factor of $\delta (1-\phi)$ when
integrated over $\omega $. The second term is more interesting. Details
of how to perform the sum over $n$ and the frequency integral may be found
in Appendix D. The final result for ${\hat {\cal P}}(\phi, x, s)$ reads
\begin{equation}
\label{ltprob}
{\hat {\cal P}}(\phi, x, s) = 
{{\hat g}(0,s)-{\hat g}(x,s) \over s{\hat g}(0,s)}
\delta (1-\phi ) \ + \ {{\hat g}(x,s)\over {\hat g}(0,s)^{2}} {1 \over
s \lambda \phi }\exp \left [ -{1 \over \lambda {\hat g}(0,s)}\log
\left ( 1\over \phi \right ) \right ] \ .
\end{equation}
This form for the Laplace transform of the probability distribution may
be easily generalized for any $d \in (0,2)$. Finally we must invert
the Laplace transform. To this end we note that
\begin{equation}
\label{ltgreen}
{\hat g}(x,s) = {1 \over (2Ds)^{1/2}} \ \exp \left [
-\left ( {2s \over D} \right )^{1/2} |x| \right ] \ .
\end{equation}
Inserting (\ref{ltgreen}) into (\ref{ltprob}) and inverting the transform
we have our final result
\begin{equation}
\label{probxt}
{\cal P}(\phi , x , t) = \delta (1-\phi) \ {\rm erf} \left [ {|x| \over
(2Dt)^{1/2} } \right ] \ + \  
{1 \over (\pi t)^{1/2}} \ {1 \over {\tilde \lambda}
\phi } \exp \left \lbrace - \left [ {|x| \over (2Dt)^{1/2}} -
{ \log (\phi ) \over 2{\tilde \lambda }t^{1/2}} \right ]^{2} \right \rbrace \ ,
\end{equation}
where ${\rm erf}(z)$ is the error function\cite{as}, and we have defined
${\tilde \lambda}=\lambda /(2D)^{1/2}$. This is illustrated in
Fig.2 for $x=1$ (and $D=1, \ {\tilde \lambda}=\sqrt{2}$), 
and three different times corresponding to
$x^{2} \gg 2Dt$, $x^{2} \sim 2Dt$ and $x^{2} \ll 2Dt$.

In particular, the probability distribution for the magnetization density
at the origin takes the form 
\begin{equation}
\label{lognor}
{\cal P}(\phi , 0 , t) = {1 \over (\pi t)^{1/2}} \ {1 \over {\tilde \lambda}
\phi } \exp \left [ -{(\log (\phi ))^{2} \over 4{\tilde \lambda }^{2}t}
\right ] \ ,
\end{equation}
which is a pure log-normal distribution. This result is very revealing, as
it shows that the fluctuations in this system are extreme. For instance,
we have already seen that the {\it mean} value of the density (at the origin) 
decays as $m(0,t) \sim 1/\sqrt{t} $. However, if one asks how the {\it typical}
(or most likely) value decays, one can see from (\ref{lognor}) that
$\langle \phi (0,t) \rangle _{\rm typ} 
\sim \exp (-{\tilde \lambda}^{2}t/2)$. Thus, as 
time proceeds, the typical value of $\phi $ decays to zero exponentially
fast, whilst the mean decays slowly as $1/\sqrt{t}$. This is possible because
the log-normal distribution has a long tail, extending out to the extreme
value of $\phi = 1$. In fact the end point of the distribution ({\it
i.e.} ${\cal P}(1,0,t)$) also decays as $1/\sqrt{t}$ which is consistent with the
known persistence properties of a random walker in $d=1$ (namely, the
probability of a walker {\it never} having returned to the origin after time
$t$ decays as $1/ \sqrt{t}$). In Fig.3 we illustrate ${\cal P}(\phi, 0 , t)$
for three different times. As a final remark, we note that if we erroneously
use the asymptotic form (\ref{resden0}) for $m(0,t;n\lambda)$ to build
the distribution function, we find that $\partial _{\phi}{\cal P}(\phi,0,t)$ is
equal to $\partial _{\phi}\delta (\phi)-\delta (1-\phi)/\lambda \sqrt{t}$;
thus emphasizing the fact that we need the entire analytic form of 
$m(0,t;n\lambda)$ to successfully construct the distribution ${\cal P}$.

This ends a rather long section on the analytic properties of the
continuum model in $d=1$. In the next section we shall briefly study
the case of higher dimensions, and then in section VI we shall compare our
results with some numerical simulations performed on the original discrete
model.

\section{Results in higher dimensions}

As we mentioned several times in the previous two sections, the
continuum theory requires some regularization for $d \ge 2$. This
can be most easily (and physically) accomplished by smearing the 
$\delta$-function interaction between the BA and the spins.
Thus, our continuum model takes the form
\begin{equation}
\label{scacontd}
\partial _{t} \phi ({\bf r},t) = -\lambda \phi ({\bf r},t)
\Delta _{l}({\bf r}-{\bf R}(t)) \ ,
\end{equation}
where $\Delta _{l}({\bf r})$ is a normalized function which is sharply
peaked over a region of linear dimension $l$ around the point ${\bf r}$.
[A good choice would be $\Delta _{l} \sim l^{-d}\exp (-r^{2}/l^{2})$.] 

In this section we shall analyze some basic properties of Eq.(\ref{scacontd})
for general $\Delta _{l}$. Then we will use a somewhat more
crude approach to estimate the mean magnetization density (at the origin)
as a function of time in $d=2$.

Although we have generalized our continuum theory somewhat, we can still
make substantial headway by first integrating the equation of motion,
and then using infinite order perturbation theory. The first step yields
\begin{equation}
\label{solcond}
\phi ({\bf r},t) = \exp \left [ -\lambda \int \limits
_{0}^{t} dt' \  \Delta _{l}({\bf r}-{\bf R}(t')) \right ] \ ,
\end{equation}
whilst the second consists of expanding this equation in powers of
$\lambda $ and averaging term by term:
\begin{equation}
\label{avdend}
m({\bf r},t) = \langle \phi ({\bf r},t) \rangle = \sum \limits _{n=0}
^{\infty} (-\lambda )^{n} \chi _{n}({\bf r},t) \ ,
\end{equation}
where $\chi _{0}({\bf r},t)=1$, and for $n>0$
\begin{equation}
\label{defchid}
\chi _{n}({\bf r},t) = {1\over n!} \left \langle \left [
\int \limits _{0}^{t} d\tau \ \Delta _{l}({\bf r}-{\bf R}(\tau)) \right ]^{n}
\right \rangle \ . 
\end{equation}
By making a Fourier representation of the interaction function $\Delta _{l}$,
performing the average over paths (see Appendix B), and finally Laplace 
transforming in time, we arrive at
\begin{equation}
\label{chid}
{\hat \chi} _{n}({\bf r},s) = {1\over s} \left \lbrace
\prod \limits _{m=1}^{n} \int {d^{d}k_{m} \over (2\pi)^{d}}
\ {{\tilde \Delta}_{l}({\bf k}_{m}-{\bf k}_{m-1}) 
\over (s+k_{m}^{2})} \right \rbrace \ \exp [i{\bf k}_{n} \cdot {\bf r}] \ ,
\end{equation}
with the convention ${\bf k}_{0} \equiv 0$.

We shall be concerned with two quantities. The first is a smeared mean
magnetization density near the origin, and the second is the global
magnetization. These are given by 
\begin{equation}
\label{defd1}
m_{\rm s}({\bf 0},t) \equiv \int d^{d}r \ \Delta _{l}({\bf r}) m({\bf r},t)
\end{equation}
and
\begin{equation}
\label{defd2}
M(t) = \int d^{d}r \ [ \langle \phi ({\bf r},0) \rangle -
\langle \phi ({\bf r},t) \rangle ]
\end{equation}
respectively. We shall prove that 
\begin{equation}
\label{idend}
{dM(t)\over dt} = \lambda m_{\rm s}({\bf 0},t) \ ,
\end{equation}
which is the smeared analogue of the global versus (strictly) local relation 
(\ref{glomagc3}) we proved for $d<2$.
Comparing Eqs.(\ref{avdend}) and (\ref{defd1}) we see that
\begin{equation}
\label{magsmex}
m_{\rm s}({\bf 0},t) = \sum \limits _{n=0}^{\infty} (-\lambda)^{n}
\eta _{n}(t) \ ,
\end{equation}
where in Laplace transform space
\begin{equation}
\label{etafns}
{\hat \eta}_{n}(s) = \int d^{d}r \ \Delta _{l}({\bf r}) 
{\hat \chi}_{n}(s) =  {1\over s} \left \lbrace
\prod \limits _{m=1}^{n} \int {d^{d}k_{m} \over (2\pi)^{d}}
\ {{\tilde \Delta}_{l}({\bf k}_{m}-{\bf k}_{m-1}) 
\over (s+k_{m}^{2})} \right \rbrace \ {\tilde \Delta} _{l}({\bf k}_{n}) \ .
\end{equation}
Similarly we have
\begin{equation}
\label{glomagsmex}
{dM(t)\over dt} = -\sum \limits _{n=0}^{\infty} (-\lambda)^{n}
\zeta _{n}(t) \ ,
\end{equation}
with
\begin{equation}
\label{relzeta}
\zeta _{n}(t) = {d \over dt} \int d^{d}r \ \chi _{n}({\bf r},t) \ .
\end{equation}
For $n>0$, $\chi _{n}({\bf r},0)=0$, so that
\begin{equation}
\label{finala}
{\tilde \zeta}_{n}(s) = s \int d^{d}r \ {\tilde \chi} _{n}({\bf r},s) \ .
\end{equation}
Using Eq.(\ref{chid}) we may evaluate the above integral to give
\begin{equation}
\label{finalb}
{\tilde \zeta}_{n}(s) =  \left \lbrace
\prod \limits _{m=1}^{n} \int {d^{d}k_{m} \over (2\pi)^{d}}
\ {{\tilde \Delta}_{l}({\bf k}_{m}-{\bf k}_{m-1}) 
\over (s+k_{m}^{2})} \right \rbrace \ \delta ^{d} ({\bf k}_{n}) \ .
\end{equation}
Performing the integral over ${\bf k}_{n}$ immediately yields
\begin{equation}
\label{finalc}
{\hat \zeta} _{n}(s) = {\tilde \eta} _{n-1}(s) \ .
\end{equation}
Thus comparing Eqs.(\ref{magsmex}), (\ref{glomagsmex}) and (\ref{finalc}) 
we see the validity of relation Eq.(\ref{idend}).

As a corollary, by integrating the averaged equation of motion 
(\ref{scacontd}) over space, we have
\begin{equation}
\label{corol1}
{dM(t) \over dt} = \lambda \int d^{d}r \ \Bigl \langle \Delta _{l}({\bf r}-
{\bf R}(t)) \phi ({\bf r},t) \Bigr \rangle \ ,
\end{equation}
When compared with Eq.(\ref{idend}), the above relation gives us 
\begin{equation}
\label{corol2}
m_{s}(t) = \int d^{d}r \ \Delta _{l}({\bf r}) \langle \phi 
({\bf r},t) \rangle = \int d^{d}r \ \langle \Delta _{l}({\bf r}-
{\bf R}(t)) \phi ({\bf r},t) \rangle \ ,
\end{equation}
which is the smeared version of the local relation (\ref{idenc}) proven in
section III. We note that, although we have been concerned with a sharply
peaked interaction function, the relations (\ref{idend}) and (\ref{corol2})
hold for {\it any} function $\Delta ({\bf r})$.

This ends the more rigorous part of the present section. In the
remainder we shall just mention some explicit results for the
mean local magnetization density (at the origin), which are obtained
with a cruder regularization.

The difficulty with making headway using the smoothing function, is
that the $n$-fold integrals over the $\Delta _{l}$'s are intractable
(unless one can find a particularly `friendly' form for $\Delta _{l}$.) 
As an alternative approach, we return to the sharp Dirac $\delta$-function
as used in section IV. We remarked that the $n$-fold convolution 
integrals were divergent due to the non-integrability of $g({\bf 0},t)$
for $d \ge 2$. To evade this difficulty we can simply impose a cut-off
into the integration limits. This is closely connected to introducing 
a microscopic time scale into the temporal correlations of the BA.
Such an regularization procedure was used in Ref. \cite{new}, and
the results so obtained were shown to be equivalent to previously
known exact results\cite{hh}. 
So we shall use the same procedure here, but with due caution.

First, we consider $d=2$. In a precisely analogous way to the calculation
in section IV, we expand the field solution (\ref{solcon}) in powers of
$\lambda $ and average term by term. Thus, we have (cf. Eqs. (\ref{avden}) -
(\ref{solchi}))
\begin{equation}
\label{avden2}
m({\bf 0},t) = \langle \phi ({\bf 0},t) \rangle = \sum \limits _{n=0}
^{\infty} (-\lambda )^{n} \chi _{n}({\bf 0},t) \ ,
\end{equation}
where $\chi _{0}({\bf 0},t)=1$, and for $n>0$
\begin{equation}
\label{solchi2}
\chi _{n}({\bf 0},t) = \int \limits _{t_{0}}^{t-t_{0}} d\tau _{1} 
\int \limits _{t_{0}}^{\tau _{1}-t_{0}} d\tau _{2} \cdots 
\int \limits _{t_{0}}^{\tau _{n-1}-t_{0}} d\tau _{n} \ 
g({\bf 0},\tau _{1}-\tau _{2}) \cdots
g({\bf 0},\tau _{n-1}-\tau _{n}) g({\bf 0},\tau _{n}) \ ,  
\end{equation}
In this case, the probability distribution of
the BA at the origin has the form $g({\bf 0},t) = (2\pi D t)^{-1}$.
Note we have inserted the microscopic time regulator $t_{0}$ in the 
limits of the time integrals. Our strategy is to evaluate the
time integrals one by one, keeping only the most singular term
at each step. We shall use the general result (for $t \gg t_{0}$)
\begin{equation}
\label{logrule}
\int \limits _{t_{0}}^{t-t_{0}} d\tau \ {[\log (\tau/t_{0})]^{m} \over
(t-\tau)\tau } \sim {(m+2) \over (m+1)} \ {[\log (t/t_{0})]^{m+1}
\over t} \ .
\end{equation}
Therefore we have the dominant contribution
\begin{equation}
\label{domcont}
\chi _{n}({\bf 0},t) \sim {[\log (t/t_{0})]^{n}\over (2\pi D)^{n}} \ .
\end{equation}
Inserting this result into Eq. (\ref{avden2}) and summing over $n$ we
have the asymptotic form
\begin{equation}
\label{avdenasy}
m({\bf 0},t) \sim \left [ 1 + {\lambda \over 2\pi D}\log \left
( {t\over t_{0}} \right ) \right ]^{-1} \ .
\end{equation}
Thus the magnetization at the origin does decay to zero for large times,
but logarithmically slowly. From the relation (\ref{idend}) we see that
the mean global magnetization increases as $M(t) \sim t/\log(t)$.

The same kind of analysis can be repeated for $d>2$, and one finds
that $m({\bf 0},t)$ saturates to a constant for large times, which
implies that $M(t) \sim t$ for large times. These results are easily
understood from the recurrent properties of the BA ({\it i.e.}
a random walker returns to its starting point with probability
one, only for $d \le 2$). It would be
more interesting to derive the distribution of the magnetization
density for $d\ge 2$, but this requires the more careful regularization
method involving the smoothing function $\Delta _{l}({\bf r})$ and
thus lies beyond the scope of the present work. 

[In $d=2$ one
can capitalize on the slightly crude result (\ref{avdenasy}) obtained
using the cut-off $t_{0}$, combined with the exact property
$\langle \phi ({\bf 0},t;\lambda )^{n} \rangle = 
m({\bf 0},t;n\lambda )$ to derive a form for the
distribution function ${\cal P}(\phi, {\bf 0}, t)$. Such an approach yields
${\cal P}(\phi, {\bf 0}, t) = \beta(t) \phi ^{\beta (t)-1}$, with 
$\beta = 2\pi D/\lambda \log (t/t_{0})$. However, this result is
not to be taken seriously, since we need the whole analytic structure
of $m({\bf 0},t;n\lambda )$ in order to derive $P$, as evidenced
in the previous section.]

\section{Numerical simulation}

We have performed extensive numerical simulations of the discrete
model, as defined in section II, in order to test the results
obtained in the last two sections from the continuum theory.
In all of the simulations for which we present results, we have set the 
hopping rate $p$ of the BA, along with the flipping probability $q$, to unity. 
We have experimented with decreasing the flipping probability, and have
found that its only effect is to renormalize the effective spin-BA
coupling $\lambda $, such that $\lambda \propto q$, as expected.

Most of our results are obtained from a one-dimensional chain of sites.
The chain length is unimportant, so long as one ensures that the
BA has never touched the edges in any of its realizations up to the
latest time at which data is extracted. Generally we average over
between $10^{6}$ and $10^{8}$ realizations (or runs) depending on the
desired quality of the data. Such simulations required a few days on
a DEC Alpha 233 MHz workstation. In a given run, at each time step the BA is moved 
left or right with equal probability and the spin it leaves behind is flipped. 
Each run starts with the same initial configuration; 
namely all spins up, except the spin at the origin 
(which is the starting site of the BA) which is 
pointed down. (This means that all spins are up after one time step, since
the BA has moved away from the origin and flipped the down spin.)

In Fig.4 we show the measured mean magnetization density at the origin,
along with the quantity $\sum _{R}\Theta (R,t,R)$ (cf. Eq.(\ref{lomag})).
They are seen to be identical thus confirming relations Eq.(\ref{iden}) and
its continuum counterpart (\ref{idenc}). The solid line is the asymptotic
prediction (\ref{resden0}) from the continuum theory. It is seen to be in 
good agreement
with the data, as the line has a slope of (-1/2). From the fit of this 
log-log plot we can read off the effective value of $\lambda $, since 
from (\ref{resden0}) the prefactor of $1/\sqrt{t}$ is given by 
$(2/\pi \lambda^{2})^{1/2}$. (The diffusion constant for the lattice random
walk is unity.) We have fitted the data to $c/\sqrt{t}$ with $c=0.40(1)$, which
yields $\lambda = 1.99(1)$. In Fig.5 we plot the small $x/\sqrt{t}$ dependence
of $m(x,t)$ on a log-log scale. The data is well fitted by the prediction
given in Eq.(\ref{resdensx1}). In Fig.6 we plot $1-m(x,t)$ versus
$x/\sqrt{t}$. Note that good data collapse is found for intermediate values
of $x/\sqrt{t}$. We have been unable to numerically probe the ballistic
scale $\lambda t$. (Note, the theoretical curves shown in the last
two figures are plotted with no free parameters.)

In Fig.7 we plot the discrete time derivative of the total number of
down spins $N(t)$ (which is $M(t)/2$), along with $m(0,t)$. The two curves
are indistinguishable within the numerical noise, thus confirming
the global/local relation (\ref{glomag3}). This also provides secondary 
confirmation of the continuum form of this relation Eq.(\ref{glomagc3}) with
$\lambda \simeq 2$.

In Fig.8 we plot the ratio of the measured two point correlation function
(divided by $m(0,t)$) versus $x/\sqrt{t}$. Note that it varies from
1/2 (at small $x$) to unity (at large $x$) as expected. The data from two
different times is shown, and one sees excellent agreement with the
theoretical prediction (\ref{corrfinal}), which 
is plotted with {\it no} free parameters. 
This agreement provides very strong evidence for the validity of our whole
continuum approach.

Briefly, we mention simulations in $d=2$. Higher dimensional simulations
are not too difficult as one is only ever moving the single BA at each
time step. In Fig.9, we show $m_{\bf 0}(t)$ and 
$\sum _{{\bf R}}\Theta ({\bf R},t,{\bf R})$. The data for the two functions
are identical verifying the discrete relation (\ref{iden}) in two dimensions,
as well as confirming the continuum result (\ref{corol2}). We have plotted 
the inverse of these functions
against $\log (t)$ in order to compare with the theoretical prediction
(\ref{avdenasy}). Again, good agreement is found, thereby confirming the less
rigorous method by which the two dimensional result was obtained.

Finally we mention our attempt to measure the probability distribution
of the coarse grained magnetization density at the origin (in $d=1$), which 
was found from the continuum theory to be a log-normal distribution. 
Clearly it does not make sense to measure moments of the spin at the 
origin, since the odd (even) moments are equal to $m_{0}(t)$ (unity).
Therefore, we define a coarse grained magnetization over a patch of 
spins. If the patch is taken too small, the coarse-graining will be 
ineffective, whilst if the patch is taken too large, the BA will take
a long time to leave the patch, and the asymptotic behavior will be
numerically inaccessible. So we have compromised and have used a patch
containing 21 spins. We have binned the patch magnetization from $10^{8}$
independent runs and generated the histograms shown in Fig.10. Note, that
because the patch size is modest, the histograms have non-zero weight
in the negative $\phi $ region, in contrast to the strict continuum
limit. However, we do see that for $\phi $ near unity, the histograms
have a robust tail, which is the signature that extreme fluctuations are
important. 

\section{Conclusions}

In this paper we have introduced and analyzed a simple model of data
corruption due to a Brownian agent (BA). In section II we introduced
a discrete version of the model, which consists of a BA
flipping bits (or spins) on a lattice. The model is non-trivial since the value
of a given spin depends very sensitively on the path of the BA ({\it
i.e.} whether the spin has been visited an odd or even number of times).
We presented a master equation formulation of the model and derived an
equation of motion for the conditional average of the magnetization
density. In the continuum limit, this quantity was seen to satisfy an
imaginary-time Schr\"odinger equation (ITSE) for a particle in a repulsive
$\delta $-function potential. Higher-order conditional averages 
also satisfy ITSE's with an additional repulsive $\delta$-function
potential for each spin being averaged. We also proved that there is an
exact proportionality between the rate of change of the 
mean global magnetization $M(t)$ (as defined
in Eq.(\ref{glomag})) and the mean magnetization density at the origin, thus
revealing a non-trivial statistical relation between a global and a local
quantity.

In section III we recast the discrete model as a stochastic cellular
automaton (SCA). From the local rules of the SCA, we postulated a 
particularly simple continuum theory written in terms of a coarse-grained
magnetization density $\phi ({\bf r},t)$ (which is a functional of the
walk ${\bf R}(t)$ of the BA). We showed that averages of this stochastic
field may be recast as integrals over the final BA position, the integrands
of which are imaginary-time path integrals (ITPI's). 
These ITPI's were shown to describe quantum systems of repulsive 
$\delta $-function potentials identical to those revealed through the
ITSE formulation of the master equation. In this way we have confirmed
that the continuum theory is a good representation of the underlying
discrete model. [It is worth mentioning that averages of $1/\phi $ 
correspond to the quantum mechanics of an {\it attractive} $\delta$-function
potential. In this case there is the possibility of a sharp transition
in behavior for $d>2$ as one varies $\lambda$, since bound states only exist 
for a sufficiently attractive well.]

In section IV we thoroughly examined the properties of the continuum
theory for $d=1$. First, we derived an exact expression for the
evolution of the magnetization density $m(x,t)$, and found its
spatial variation for small and large $x$. In particular we found 
$m(0,t) \sim 1/\lambda \sqrt{t}$ for large times; and also the existence of
both a diffusive and a ballistic scale in $m(x,t)$. Using the exact result 
for $m(x,t)$ enabled us to prove that the rate of change of $M(t)$ is
proportional to $m(0,t)$ for all times (with proportionality constant
$\lambda $) which is the continuum analog of the global/local relation
that was proved in section II. We then studied the two-point correlation
function $C(x,t)$. An exact expression was found for the Laplace transform
of this quantity, from which we were able to extract its long-time behavior,
which is expressed in terms of a Jacobi theta function as shown in 
Eq.(\ref{corrfinal}). The correlation function 
has an asymptotic scaling form; namely the
ratio $C(x,t)/m(0,t)$ depends only on $x/\sqrt{t}$. Finally in
section IV, we examined the probability density ${\cal P}$ of the magnetization
density. Using a fortuitous property of the original continuum theory which
enables us to represent the $n^{\rm th}$ moment of the magnetization
density in terms of $m(x,t)$, along with our exact expression for this
latter quantity, we were able to derive an exact expression for 
${\cal P}(\phi, x, t)$ as shown in Eq.(\ref{probxt}). In particular, setting $x=0$
reveals that ${\cal P}(\phi, 0, t)$ is a pure log-normal distribution. This last
result emphasizes the importance of extreme fluctuations in this system.
For instance, $m(0,t)$ decays as $1/\sqrt{t}$, whilst the typical value
of the magnetization density at the origin (i.e. the mode of ${\cal P}$) decays
as $\exp (-\lambda ^{2}t/4D)$.

In section V we briefly examined higher dimensions. For $d \ge 2$ it
is necessary to regularize the continuum theory, most appropriately
with a smeared interaction between the BA and the spins. Using an
arbitrary interaction function $\Delta _{l}({\bf r})$ (which has
a linear scale $l$) we were able to prove a more general form of
the global/local relation, as shown in Eq.(\ref{idend}). We then concentrated
on $d=2$, and using a cruder regularization (namely, introducing a 
microscopic correlation time $t_{0}$) we were able to derive an expression
for the asymptotic decay of $m({\bf 0},t)$. This decay 
(cf. Eq.(\ref{avdenasy})) is seen to be logarithmically slow.

In section VI we presented our results from extensive numerical simulations
of the original lattice model. Most of our numerical work examines
the case of $d=1$. We measured the spatial and temporal variation of
the mean magnetization density, the mean global magnetization, and the
two-point correlation function $C_{x}(t)$. 
In all cases we found excellent agreement
between our data and the theoretical predictions arising from the 
continuum model. In particular, the agreement between the theoretical
form for $C(x,t)$ and the numerical data is very satisfying, as there
are no free parameters to adjust. We also measured $m_{\bf 0}(t)$ in
two dimensional simulations, and found good agreement with the
predicted logarithmic decay. Finally we attempted to measure the 
probability distribution of the magnetization density, by binning
the magnetization of a spin patch (containing 21 spins) from $10^{8}$
realizations. The results are not of high enough quality to directly
compare with the derived log-normal distribution; however, we do see
clear evidence of a long robust tail of the histogram for values of
$\phi $ near unity, which is a clear signature of the importance of
extreme fluctuations.

In conclusion, we have introduced and solved a model in which a BA
interacts with a bimodal environment ({\it i.e.} a medium
containing two types of particles, spins, bits, etc.). Our primary
application has been an environment composed of bits of data, which 
the BA steadily corrupts. We have been interested in the statistical
correlations and fluctuations of the disordered medium, and our exact results
(arising from a simple continuum model) reveal the correlations to
be non-trivial and the fluctuations to be extreme in nature. These
statements are made quantitative by the form of the two-point correlation
function, and the probability distribution of the density of corrupted bits.
There are many directions for future work, foremost amongst which are: i)
calculating the distribution ${\cal P}$ 
in $d=2$ and determining its sensitivity
to the smearing function $\Delta _{l}$, ii) investigating autocorrelation
effects in one dimension, iii) refining the model to take into account
relevant factors (like a back-coupling between the environment and the BA,
or an asymmetry in the flipping probability) which will arise if
one tries to make a stronger connection to real processes.

The model may also be seen to be a very simplified version of other
systems. For instance the bistable medium can be taken
to be composed of two chemical species A and B (with vanishingly low mobility)
and the BA to be a high mobility catalyst, inducing a reversible
reaction between A and B (and {\it vice versa}). Alternatively we can
think of the BA as a wandering impurity in an ionic crystal
(such as an anion or cation vacancy in {\it NaCl}) or a semiconductor 
compound (such as {\it Zn} in {\it GaAs}),
which has a small probability of reordering the local bi-atomic structure
as it passes through a given unit cell\cite{al}. 

The data corruption process
appears to us the most interesting application, as well has being the
most potentially relevant. This is especially true given the enormous
efforts dedicated to creating memory storage devices of ever-decreasing size. 
Such miniaturization will lead to new causes of soft error 
production\cite{shar}, 
amongst which will inevitably be found the Brownian agent.

The authors would like to thank R. Burghaus, M. Howard, W. Loinaz, 
A. McKane, B. Schmittmann and R. Zia for useful discussions. 
The authors also gratefully acknowledge financial support from the 
Division of Materials Research of the National Science Foundation. 

\newpage

\appendix

\section{}

In this appendix we outline the solution to the discrete equation
(\ref{lomageq}) for the conditional average 
$\Theta ({\bf r},t \ | \ {\bf R})$.
This is achieved by use of discrete Fourier and Laplace transforms
defined via 
\begin{equation}
\label{app1_1}
{\cal F}_{k|R} [ {\cal L}_{z|t} [ \Theta  ({\bf r},t \ | \ {\bf R})]] = 
\sum \limits _{n=0}^{\infty} z^{n} \ \sum \limits _{\bf R}
 \Theta ({\bf r},n \delta t\ | \ {\bf R}) \exp (i{\bf k} \cdot {\bf R}) \ ,
\end{equation}
 (with $t=n\delta t$).
By self-consistently determining the function $\Theta ({\bf r},t \ | \ {\bf r})$,
one can explicitly solve for the double transform of $\Theta$. Inverting the
discrete Fourier transform and summing over ${\bf R}$ one finds
(with the particular choice of initial condition (\ref{ict}))
\begin{equation}
\label{app1_2}
\sum \limits _{\bf R} {\cal L}_{z|t} [\Theta  ({\bf R}, t \ | \ {\bf R})] = 
{1 \over (1-z)} \left \lbrace {1 - 2(1-z) \int \limits _{\rm BZ}
d^{d}k \ E({\bf k},z) \over 1 + {zq \over d}  \int \limits _{\rm BZ}
d^{d}k \ E({\bf k},z) \ \sum _{\bf l} e^{-i{\bf k}\cdot {\bf l}}}
\right \rbrace \ ,
\end{equation}
where $\int _{\rm BZ}$ indicates integration over the first Briulloin
zone, and 
\begin{equation}
\label{app1_3}
E({\bf k},z) = \left \lbrace 1 - z + {qz \over 2d} \sum \limits _{\bf l}
[ 1 - \exp (-i{\bf k} \cdot {\bf l}) ] \right \rbrace ^{-1} \ . 
\end{equation}
Similarly one can find an explicit expression for 
$\sum \limits _{\bf R} {\cal L}_{z|t} [\Theta  ({\bf 0}, t \ | \ {\bf R})]$.
The expressions can be shown to be equal thus proving
the relation (\ref{iden}) as given in the main text.

\section{}

In this appendix we outine the procedure for averaging the expression
in Eq.(\ref{defchi}). First, we time order the integrals, thereby absorbing
the factor of $1/n!$. Then, making an integral representation of each
Dirac $\delta $-function, we have
\begin{eqnarray}
\label{app2_1}
\nonumber
\chi _{n}(x,t) & = & \int \limits _{0}^{t} d\tau _{1} 
\int \limits _{0}^{\tau _{1}} d\tau _{2} \cdots 
\int \limits _{0}^{\tau _{n-1}} d\tau _{n}
\int dk_{1}e^{ik_{1}x} \cdots \int dk_{n}e^{ik_{n}x}\\
& \times & \Bigl \langle \exp \left [
-ik_{1}R(\tau _{1}) - \cdots -ik_{n}R(\tau _{n}) \right ]
\Bigr \rangle \ .
\end{eqnarray}
Using the solution of Eq.(\ref{scacont1}), we may rewrite the
average in (\ref{app2_1}) as
\begin{equation}
\label{app2_2}
\left \langle \exp \left [
-ik_{1}\int \limits _{\tau _{2}}^{\tau _{1}} d\tau \xi (\tau )
-i(k_{1}+k_{2})\int \limits _{\tau _{3}}^{\tau _{2}} d\tau \xi (\tau )
- \cdots
-i(k_{1}+ \cdots + k_{n})\int \limits _{0}^{\tau _{n}} d\tau \xi (\tau )
\right ] \right \rangle \ .
\end{equation}
The Gaussian average over the noise may now be performed, and the above 
expression reduces to
\begin{equation}
\label{app2_3}
\exp \left \lbrace -{D \over 2} \bigl [ k_{1}^{2}(\tau _{1}-\tau _{2})
+ (k_{1}+k_{2})^{2}(\tau _{2}-\tau _{3}) + \cdots
+ (k_{1}+\cdots +k_{n})^{2}\tau _{n} \bigr ] \right \rbrace \ . 
\end{equation}
We now insert this expression back into Eq.(\ref{app2_1}), and make the
change of variables $k_{1}'=k_{1}, \ k_{2}'=k_{1}+k_{2}, \cdots
, k_{n}'=k_{1}+\cdots +k_{n}$. The integrals over $\lbrace k_{j}' \rbrace$
are easily performed and we arrive at Eq.(\ref{solchi}) in the main text.

\section{}

In this appendix we outline the evaluation of the $2n$-fold integral for 
$c_{n}(x,t)$ given in Eq.(\ref{corrcn}). We note first that the time 
integrals have the form of a $n$-fold convolution. Thus, we can
Laplace transform the expression to find 
\begin{eqnarray}
\label{app3_1}
\nonumber
{\hat c}_{n}(x,s) & = & {1 \over s} \ 
\int {dk_{1}\over 2\pi} \ (1+e^{-ik_{1}x}) \cdots
\int {dk_{n}\over 2\pi} \ (1+e^{-ik_{n}x}) \\
& & \ \ \ \ \ \ \ \ \ \times \prod \limits _{m=1}^{n} {1 \over \bigl 
[ s +(D/2)(k_{1}+\cdots +k_{m})^{2} \bigr ] } \ .
\end{eqnarray}
We make the change of variable $k_{1}'=k_{1}, \ k_{2}'=k_{1}+k_{2}, \cdots
, k_{n}'=k_{1}+\cdots +k_{n}$, and rearrange the integrals to give
\begin{eqnarray}
\label{app3_2}
\nonumber
{\hat c}_{n}(x,s) & = & {1 \over s} \ 
\int {dk_{1}'\over 2\pi} \ [s+(D/2){k_{1}'}^{2}]^{-1} \cdots
\int {dk_{n}'\over 2\pi} \ [s+(D/2){k_{n}'}^{2}]^{-1} \\
& & \ \ \ \ \ \ \ \ \ \ \ \ \ \times \prod \limits _{m=1}^{n}  
\Bigl [ 1 + \exp [-i (k_{m}'-k_{m-1}')x] \Bigr ]  \ ,
\end{eqnarray}
with the understanding that $k_{0}'\equiv 0$.
We now multiply out the product which gives us $2^{n}$ terms. These
are grouped into $n+1$ sets, the $m^{\rm th}$ set containing
$C^{n}_{m}$ terms which are equal after integration. Thus
each term in the $m^{\rm th}$
set (within the integrals) may be taken to contain $m$ factors of
$e^{ik_{j}x}$ (where $j=1,\cdots,m$) and
$(n-m)$ factors of unity.
Using the integral
\begin{equation}
\label{app3_3}
\int \limits _{-\infty}^{\infty}{dk \over 2\pi} \ {e^{ikx} \over
[s + (D/2)k^{2}]} = {1 \over (2sD)^{1/2}} \exp \left [
- \left ( {2s\over D} \right )^{1/2}|x| \right ] \ ,
\end{equation}
we can reduce Eq.(\ref{app3_2}) to the form
\begin{equation}
\label{app3_4}
{\hat c}_{n}(x,s) = {1 \over s} \sum \limits _{m=0}^{n}
C^{n}_{m} \left [ {1 \over (2sD)^{1/2} } \right ] ^{n}
\exp \left [- m\left ( {2s\over D} \right )^{1/2}|x| \right ] \ .
\end{equation}
Performing the binomial sum gives Eq.(\ref{ltcorrn}) in the main text.

\section{}

In this appendix we outline the derivation of Eq.(\ref{ltprob}) 
from Eqs.(\ref{deltaex}) and (\ref{ltdengen}). The only non-trivial
aspect of the derivation is the frequency integral and sum (over $n$)
of the second term in Eq.(\ref{ltdengen}). Ignoring the prefactor of that
term (namely ${\hat g}(x,s)/s{\hat g}(0,s)$) we must evaluate a quantity
$Q(s,\phi)$, which has the form
\begin{equation}
\label{app4_1}
Q(s,\phi ) \equiv 
\int \limits _{-\infty }^{\infty} {d\omega \over 2\pi} e^{-i\omega \phi}
\ \sum \limits _{n=0}^{\infty} \ {(i\omega )^{n}\over n!} 
\ {1 \over (1+n\lambda {\hat g}(0,s))} \ .
\end{equation}
In order to perform the sum, we introduce the integral representation
\begin{equation}
\label{app4_2}
{1 \over (1+n\lambda {\hat g}(0,s))} = \int \limits _{0}^{\infty} du \
\exp \Bigl [ -u(1 + n\lambda {\hat g}(0,s)) \Bigr ] \ .
\end{equation}
The sum over $n$ now reconstitutes an exponential function, and we have
\begin{equation}
\label{app4_3}
Q(s,\phi ) = \int \limits _{0}^{\infty} du \ e^{-u} \
\int \limits _{-\infty }^{\infty} {d\omega \over 2\pi} e^{-i\omega \phi}
\exp \left ( i\omega e^{-u\lambda {\hat g}} \right ) \ .
\end{equation}
The integral over $\omega $ is easily done to give
\begin{equation}
\label{app4_4}
Q(s,\phi ) = \int \limits _{0}^{\infty} du \ e^{-u} \
\delta \left ( \phi -  e^{-u\lambda {\hat g}} \right ) \ .
\end{equation}
Finally, changing variables to $v= e^{-u\lambda {\hat g}}$ we have
\begin{equation}
\label{app4_5}
Q(s,\phi ) = {1 \over \lambda {\hat g}(0,s) \phi}
\exp \left [ -{1 \over \lambda {\hat g}(0,s)}\log
\left ( 1\over \phi \right ) \right ] \ .
\end{equation}
Using this result for $Q$ in conjunction with  Eqs.(\ref{deltaex}) 
and (\ref{ltdengen}) we have (\ref{ltprob}) in the main text.

\begin{figure}[tbp]
\centerline{\epsfxsize=17.0cm
\epsfbox{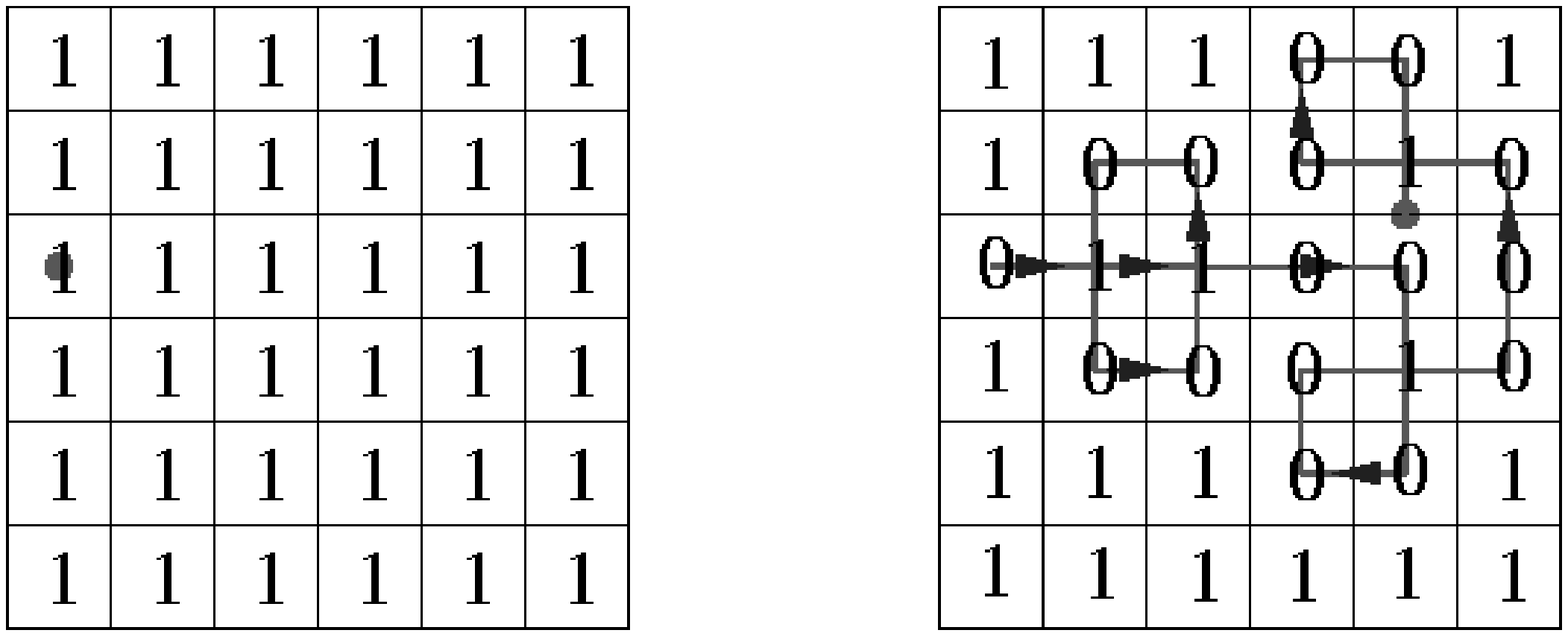}}
\vspace{0.4cm}
\caption{Illustration of the data corruption process for $d=2$. The
initial uncorrupted state is shown on the left, with the BA represented
by the filled circle. On the right we show a typical walk of $\sim 20$ steps.
The BA flips a spin with each visit, so those spins visited an even number
of times are restored to their original value.}
\end{figure}

\begin{figure}[tbp]
\centerline{\epsfxsize=10.0cm
\epsfbox{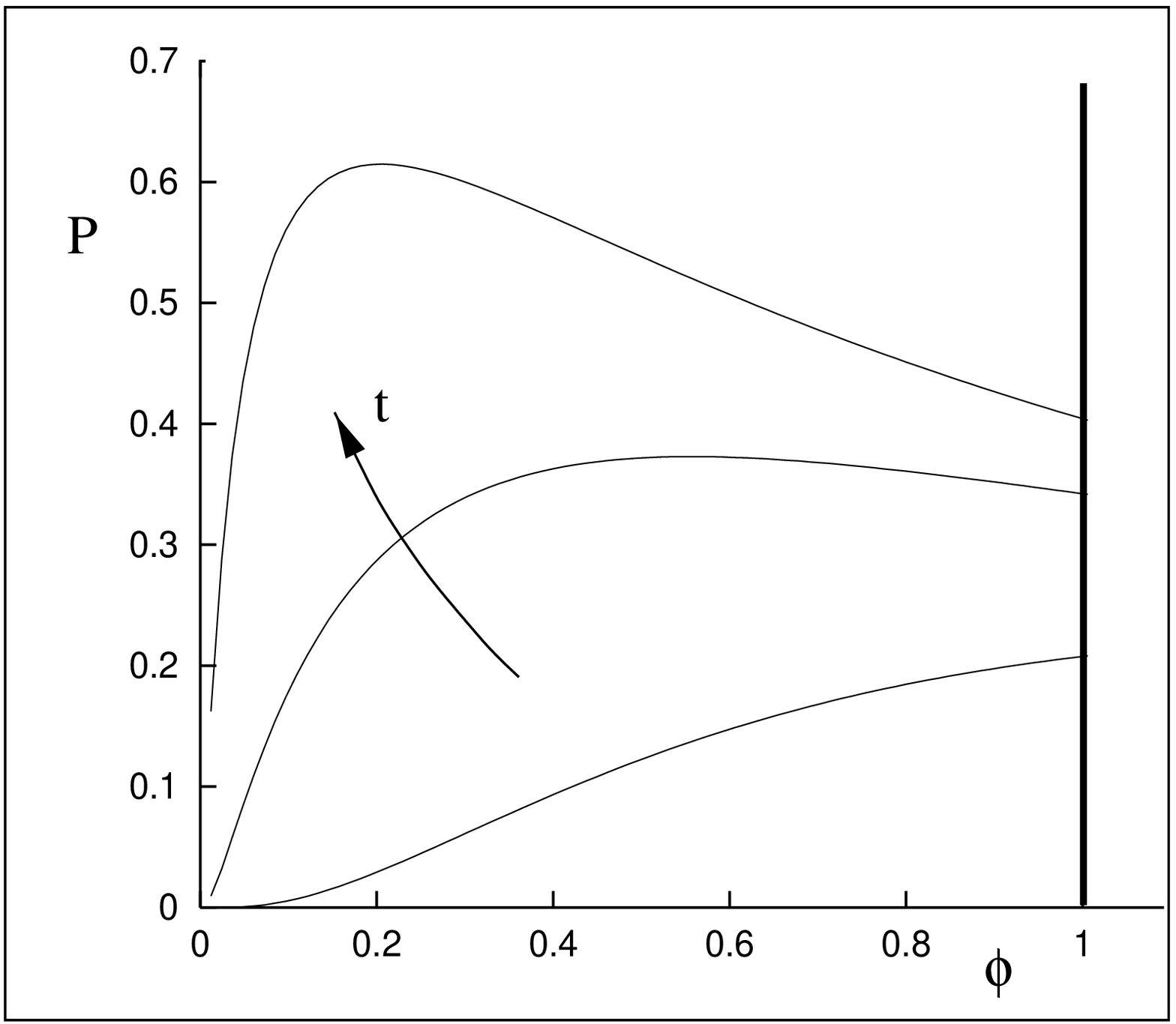}}
\vspace{0.4cm}
\caption{${\cal P}(\phi ,x,t)$ versus $\phi$, as given in Eq.(\ref{probxt}), 
with $x=1, {\tilde \lambda }=\sqrt{2}$, for three different times, 
$t=0.5, \ 1.0, \ 1.5$. The thick vertical line represents 
the $\delta$-function at $\phi =1$.} 
\end{figure}

\begin{figure}[tbp]
\centerline{\epsfxsize=10.0cm
\epsfbox{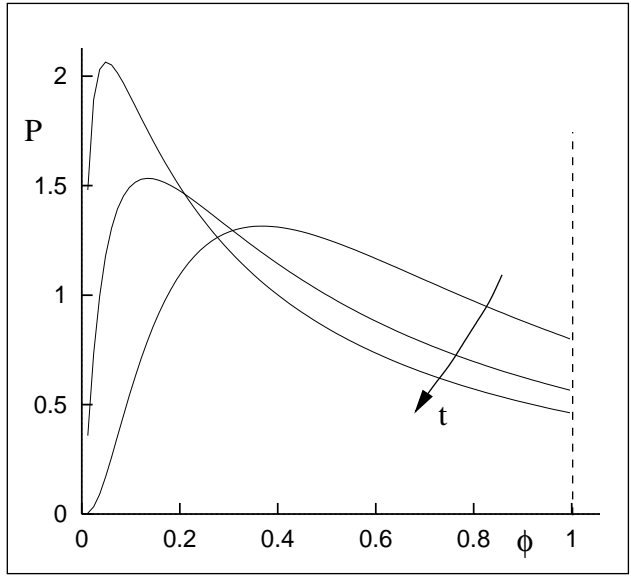}}
\vspace{0.4cm}
\caption{${\cal P}(\phi ,0,t)$ versus $\phi $, as given in Eq.(\ref{lognor}), 
for three different times.}
\end{figure}

\vspace{1.5cm}

\begin{figure}[tbp]
\centerline{\epsfxsize=10.0cm
\epsfbox{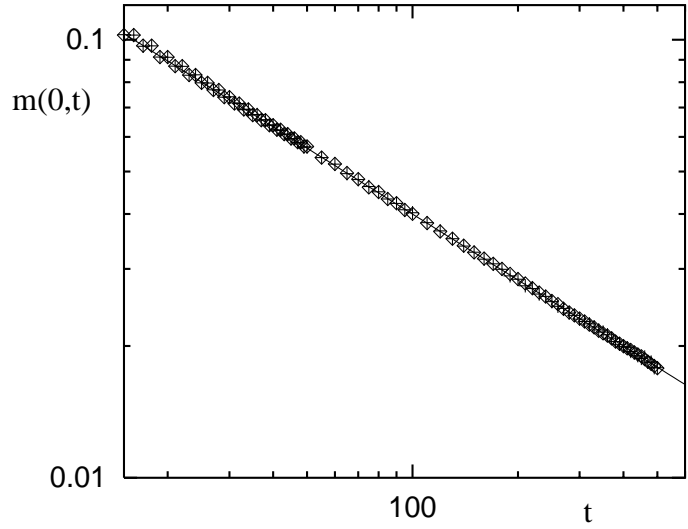}}
\vspace{0.4cm}
\caption{Log-log plot of $m_{0}(t)$ (diamonds) and $\sum _{R}\Theta (R,t,R)$ 
(plusses) versus time from $d=1$ numerical simulation. The solid line is the 
asymptotic theoretical prediction (\ref{resden0}).}
\end{figure}

\begin{figure}[tbp]
\centerline{\epsfxsize=10.0cm
\epsfbox{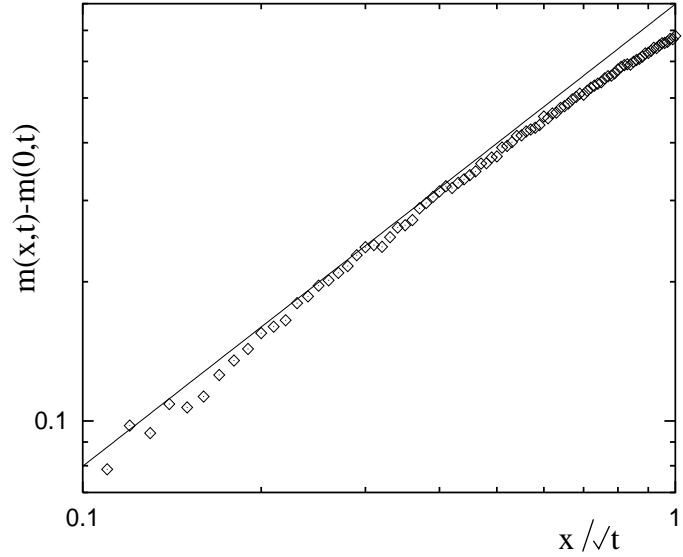}}
\vspace{0.4cm}
\caption{Log-log plot of $m_{x}(t)-m_{0}(t)$ versus $x/\sqrt{t}$ for a
time of $10^{3}$ in $d=1$. The solid line is 
the theoretical prediction (\ref{resdensx1}).}
\end{figure}

\vspace{1.5cm}

\begin{figure}[tbp]
\centerline{\epsfxsize=10.0cm
\epsfbox{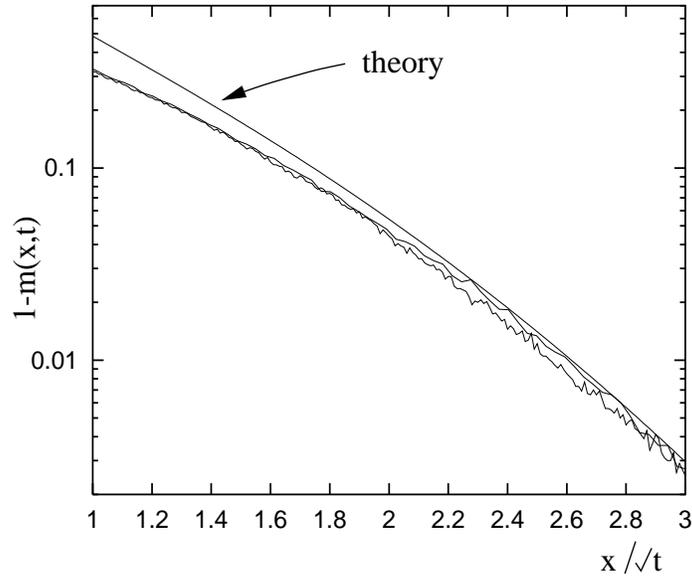}}
\vspace{0.4cm}
\caption{Log-log plot of $1-m_{0}(t)$ versus $x/\sqrt{t}$ for
times $10^{3}$ and $10^{4}$ in $d=1$. Also shown is 
the theoretical prediction (\ref{resdensx2}) for the intermediate regime.}
\end{figure}

\begin{figure}[tbp]
\centerline{\epsfxsize=10.0cm
\epsfbox{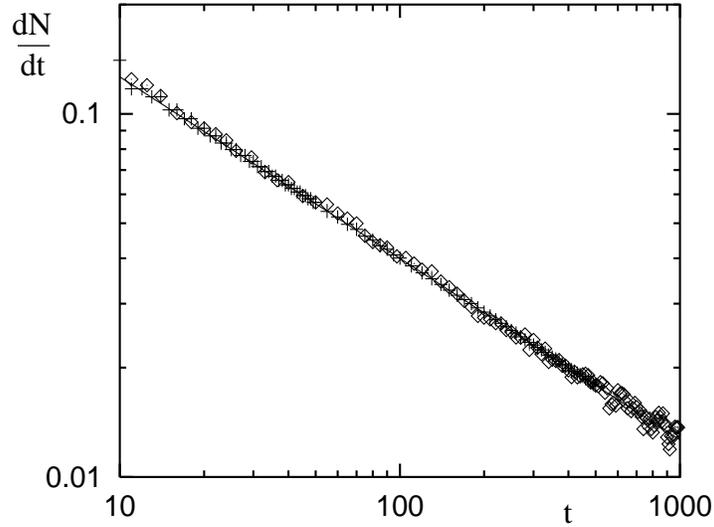}}
\vspace{0.4cm}
\caption{Log-log plot of the time derivative of the total number
of minus spins (diamonds) and of $m_{0}(t)$ (plusses) versus time
in $d=1$. The solid line is the theoretical prediction  (\ref{resden0}).}
\end{figure}

\vspace{1.5cm}

\begin{figure}[tbp]
\centerline{\epsfxsize=10.0cm
\epsfbox{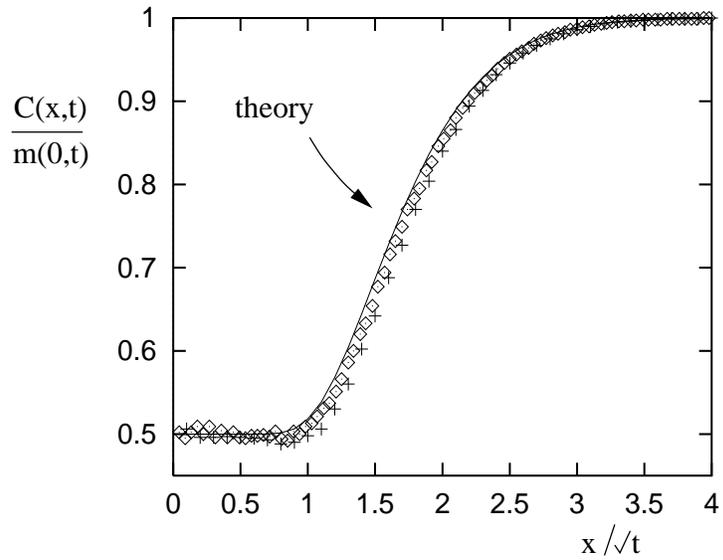}}
\vspace{0.4cm}
\caption{The ratio $C_{x}(t)/m_{0}(t)$ plotted against $x/\sqrt{t}$ for 
times of $100$ (plusses) and $500$ (diamonds) in $d=1$. The solid 
line is the theoretical prediction (\ref{corrfinal}).}
\end{figure}

\begin{figure}[tbp]
\centerline{\epsfxsize=10.0cm
\epsfbox{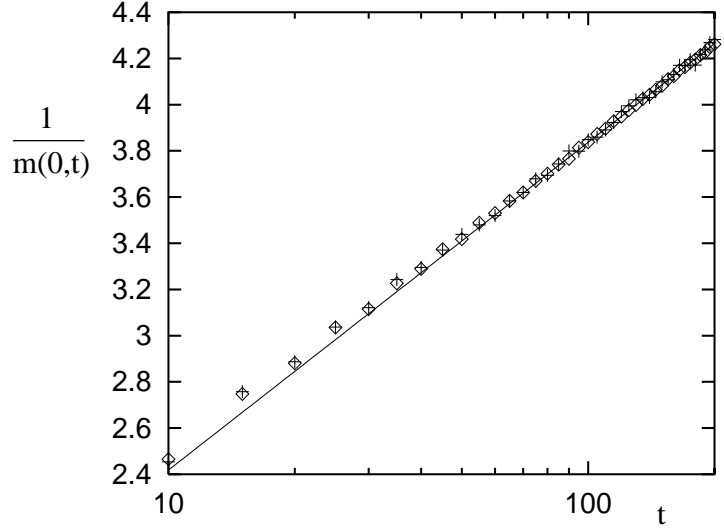}}
\vspace{0.4cm}
\caption{Linear-log plot of $1/m_{\bf 0}(t)$ versus time for $d=2$. The
solid line is the theoretical prediction (\ref{avdenasy}).}
\end{figure}

\vspace{1.5cm}

\begin{figure}[tbp]
\centerline{\epsfxsize=10.0cm
\epsfbox{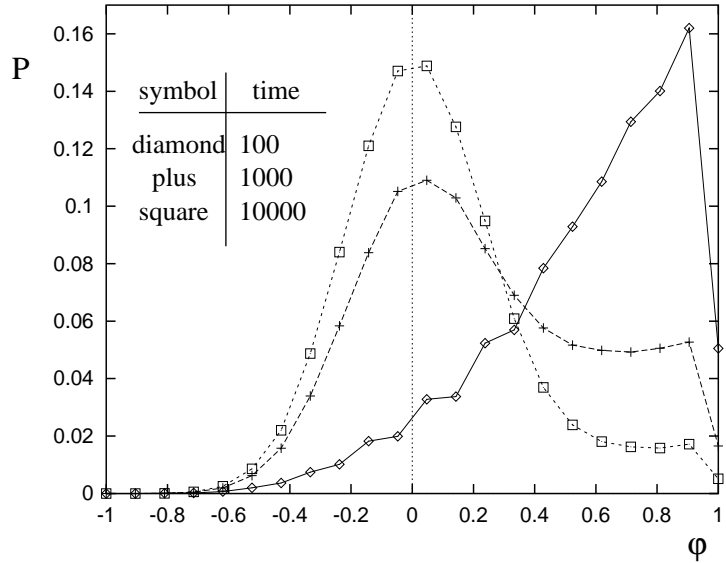}}
\vspace{0.4cm}
\caption{Probability distribution of patch (21 spins) magnetization for
times of $10^{2}$, $10^{3}$, and $10^{4}$ in $d=1$. Note the robust tail for
values of patch magnetization near unity.}
\end{figure}

\end{document}